\documentclass[aps,prd,showpacs,nobibnotes,nofootinbib,twocolumn,preprintnumbers,nobalancelastpage,superscriptaddress]{revtex4}
\usepackage{xspace}
\usepackage{subfigure}
\usepackage{graphics}
\usepackage{latexsym}
\usepackage{dcolumn}
\usepackage{bm}
\usepackage{amsmath}
\usepackage{multirow}
\usepackage{rotating}
\input{epsf}
\newcommand{\be}{\begin{equation}}
\newcommand{\ee}{\end{equation}}
\newcommand{\ba}{\begin{eqnarray}}
\newcommand{\ea}{\end{eqnarray}}
\newcommand{\no}{\no
}
\newcommand{\bfi}{\begin{figure}[!htbp]
\epsfxsize=9cm
\epsffile}
\newcommand{\efi}{\end{figure}}
\newcommand{\bi}{\begin{itemize}}
\newcommand{\ei}{\end{itemize}}

\renewcommand{\deg}{^{\circ}}
\newcommand{\lsa}{\textsc{LSA}\xspace}
\newcommand{\msa}{\textsc{MSA}\xspace}

\newcommand{\dec}{\textsc{dec.}\xspace}
\newcommand{\argo}{\textsc{ARGO-YBJ}\xspace}
\newcommand{\ra}{\textsc{r.a.}\xspace}

\newcommand{\apjl}{Astrophys. J. Lett. }

\newcommand{\nima}{Nucl. Instrum. Methods Phys. Res., Sect. A }
\newcommand{\etal}{\emph{et al.}}

\usepackage{epsfig}

\begin{document}

\date{\today}
\title{Medium scale anisotropy in the TeV cosmic ray flux observed by ARGO-YBJ}

\author{B.~Bartoli}
 \affiliation{Dipartimento di Fisica dell'Universit\`a di Napoli
                  ``Federico II'', Complesso Universitario di Monte
                  Sant'Angelo, via Cinthia, 80126 Napoli, Italy.}
 \affiliation{Istituto Nazionale di Fisica Nucleare, Sezione di
                  Napoli, Complesso Universitario di Monte
                  Sant'Angelo, via Cinthia, 80126 Napoli, Italy.}
\author{P.~Bernardini}
 \affiliation{Dipartimento Matematica e Fisica "Ennio De Giorgi", 
                  Universit\`a del Salento, via per Arnesano, 73100 Lecce, Italy.}
 \affiliation{Istituto Nazionale di Fisica Nucleare, Sezione di
                  Lecce, via per Arnesano, 73100 Lecce, Italy.}
\author{X.J.~Bi}
 \affiliation{Key Laboratory of Particle Astrophysics, Institute 
                  of High Energy Physics, Chinese Academy of Sciences,
                  P.O. Box 918, 100049 Beijing, P.R. China.}
\author{I.~Bolognino}
 \affiliation{Dipartimento di Fisica dell'Universit\`a di Pavia, via Bassi 6,
                  27100 Pavia, Italy.}
 \affiliation{Istituto Nazionale di Fisica Nucleare, Sezione di Pavia, 
                  via Bassi 6, 27100 Pavia, Italy.}
\author{P.~Branchini}
  \affiliation{Istituto Nazionale di Fisica Nucleare, Sezione di
                  Roma Tre, via della Vasca Navale 84, 00146 Roma, Italy.}
\author{A.~Budano}
 \affiliation{Istituto Nazionale di Fisica Nucleare, Sezione di
                  Roma Tre, via della Vasca Navale 84, 00146 Roma, Italy.}
\author{A.K.~Calabrese Melcarne}
 \affiliation{Istituto Nazionale di Fisica Nucleare - CNAF, Viale 
                  Berti-Pichat 6/2, 40127 Bologna, Italy.}
\author{P.~Camarri}
 \affiliation{Dipartimento di Fisica dell'Universit\`a di Roma ``Tor  Vergata'', via della Ricerca Scientifica 1, 00133 Roma, Italy.}
 \affiliation{Istituto Nazionale di Fisica Nucleare, Sezione di
                   Roma Tor Vergata, via della Ricerca Scientifica 1, 
                   00133 Roma, Italy.}
\author{Z.~Cao}
 \affiliation{Key Laboratory of Particle Astrophysics, Institute 
                  of High Energy Physics, Chinese Academy of Sciences,
                  P.O. Box 918, 100049 Beijing, P.R. China.}
 \author{R.~Cardarelli}
 \affiliation{Istituto Nazionale di Fisica Nucleare, Sezione di
                   Roma Tor Vergata, via della Ricerca Scientifica 1, 
                   00133 Roma, Italy.}
 \author{S.~Catalanotti}
 \affiliation{Dipartimento di Fisica dell'Universit\`a di Napoli
                  ``Federico II'', Complesso Universitario di Monte 
                  Sant'Angelo, via Cinthia, 80126 Napoli, Italy.}
 \affiliation{Istituto Nazionale di Fisica Nucleare, Sezione di
                  Napoli, Complesso Universitario di Monte
                  Sant'Angelo, via Cinthia, 80126 Napoli, Italy.}
 \author{S.Z.~Chen}
 \affiliation{Key Laboratory of Particle Astrophysics, Institute 
                  of High Energy Physics, Chinese Academy of Sciences,
                  P.O. Box 918, 100049 Beijing, P.R. China.}
 \author{T.L.~Chen}
 \affiliation{Tibet University, 850000 Lhasa, Xizang, P.R. China.}
 \author{P.~Creti}
 \affiliation{Istituto Nazionale di Fisica Nucleare, Sezione di
                  Lecce, via per Arnesano, 73100 Lecce, Italy.}
 \author{S.W.~Cui}
 \affiliation{Hebei Normal University, Shijiazhuang 050016, 
                   Hebei, P.R. China.}
 \author{B.Z.~Dai}
 \affiliation{Yunnan University, 2 North Cuihu Rd., 650091 Kunming, 
                   Yunnan, P.R. China.}
 \author{A. D'Amone}
 \affiliation{Dipartimento Matematica e Fisica "Ennio De Giorgi", 
                  Universit\`a del Salento, via per Arnesano, 73100 Lecce, Italy.}
 \affiliation{Istituto Nazionale di Fisica Nucleare, Sezione di
                  Lecce, via per Arnesano, 73100 Lecce, Italy.}
 \author{Danzengluobu}
 \affiliation{Tibet University, 850000 Lhasa, Xizang, P.R. China.}
 \author{I.~De Mitri}
 \affiliation{Dipartimento Matematica e Fisica "Ennio De Giorgi", 
                  Universit\`a del Salento, via per Arnesano, 73100 Lecce, Italy.}
  \affiliation{Istituto Nazionale di Fisica Nucleare, Sezione di
                  Lecce, via per Arnesano, 73100 Lecce, Italy.}
 \author{B.~D'Ettorre Piazzoli}
 \affiliation{Dipartimento di Fisica dell'Universit\`a di Napoli
                  ``Federico II'', Complesso Universitario di Monte 
                  Sant'Angelo, via Cinthia, 80126 Napoli, Italy.}
 \affiliation{Istituto Nazionale di Fisica Nucleare, Sezione di
                  Napoli, Complesso Universitario di Monte
                  Sant'Angelo, via Cinthia, 80126 Napoli, Italy.}
 \author{T.~Di Girolamo}
 \affiliation{Dipartimento di Fisica dell'Universit\`a di Napoli
                  ``Federico II'', Complesso Universitario di Monte 
                  Sant'Angelo, via Cinthia, 80126 Napoli, Italy.}
 \affiliation{Istituto Nazionale di Fisica Nucleare, Sezione di
                  Napoli, Complesso Universitario di Monte
                  Sant'Angelo, via Cinthia, 80126 Napoli, Italy.}
 \author{ G.~Di Sciascio}
\altaffiliation{Corresponding author: giuseppe.disciascio@roma2.infn.it}
 \affiliation{Istituto Nazionale di Fisica Nucleare, Sezione di
                   Roma Tor Vergata, via della Ricerca Scientifica 1, 
                   00133 Roma, Italy.}
 \author{C.F.~Feng}
 \affiliation{Shandong University, 250100 Jinan, Shandong, P.R. China.}
 \author{Zhaoyang Feng}
 \affiliation{Key Laboratory of Particle Astrophysics, Institute 
                  of High Energy Physics, Chinese Academy of Sciences,
                  P.O. Box 918, 100049 Beijing, P.R. China.}
 \author{Zhenyong Feng}
 \affiliation{Southwest Jiaotong University, 610031 Chengdu, 
                   Sichuan, P.R. China.}
\author{Q.B.~Gou}
 \affiliation{Key Laboratory of Particle Astrophysics, Institute 
                  of High Energy Physics, Chinese Academy of Sciences,
                  P.O. Box 918, 100049 Beijing, P.R. China.}
 \author{Y.Q.~Guo}
 \affiliation{Key Laboratory of Particle Astrophysics, Institute 
                  of High Energy Physics, Chinese Academy of Sciences,
                  P.O. Box 918, 100049 Beijing, P.R. China.}
 \author{H.H.~He}
 \affiliation{Key Laboratory of Particle Astrophysics, Institute 
                  of High Energy Physics, Chinese Academy of Sciences,
                  P.O. Box 918, 100049 Beijing, P.R. China.}
 \author{Haibing Hu}
 \affiliation{Tibet University, 850000 Lhasa, Xizang, P.R. China.}
 \author{Hongbo Hu}
 \affiliation{Key Laboratory of Particle Astrophysics, Institute 
                  of High Energy Physics, Chinese Academy of Sciences,
                  P.O. Box 918, 100049 Beijing, P.R. China.}
 \author{M.~Iacovacci}
 \affiliation{Dipartimento di Fisica dell'Universit\`a di Napoli
                  ``Federico II'', Complesso Universitario di Monte 
                  Sant'Angelo, via Cinthia, 80126 Napoli, Italy.}
 \affiliation{Istituto Nazionale di Fisica Nucleare, Sezione di
                  Napoli, Complesso Universitario di Monte
                  Sant'Angelo, via Cinthia, 80126 Napoli, Italy.}
 \author{R.~Iuppa}
\altaffiliation{Corresponding author: roberto.iuppa@roma2.infn.it}
 \affiliation{Dipartimento di Fisica dell'Universit\`a di Roma ``Tor Vergata'', via della Ricerca Scientifica 1, 00133 Roma, Italy.}
 \affiliation{Istituto Nazionale di Fisica Nucleare, Sezione di
                   Roma Tor Vergata, via della Ricerca Scientifica 1, 
                   00133 Roma, Italy.}
 \author{H.Y.~Jia}
 \affiliation{Southwest Jiaotong University, 610031 Chengdu, 
                   Sichuan, P.R. China.}
 \author{Labaciren}
 \affiliation{Tibet University, 850000 Lhasa, Xizang, P.R. China.}
 \author{H.J.~Li}
 \affiliation{Tibet University, 850000 Lhasa, Xizang, P.R. China.}
\author{G.~Liguori}
 \affiliation{Dipartimento di Fisica dell'Universit\`a di Pavia, via Bassi 6,
                  27100 Pavia, Italy.}
 \affiliation{Istituto Nazionale di Fisica Nucleare, Sezione di Pavia, 
                  via Bassi 6, 27100 Pavia, Italy.}
 \author{C.~Liu}
 \affiliation{Key Laboratory of Particle Astrophysics, Institute 
                  of High Energy Physics, Chinese Academy of Sciences,
                  P.O. Box 918, 100049 Beijing, P.R. China.}
 \author{J.~Liu}
 \affiliation{Yunnan University, 2 North Cuihu Rd., 650091 Kunming, 
                   Yunnan, P.R. China.}
 \author{M.Y.~Liu}
 \affiliation{Tibet University, 850000 Lhasa, Xizang, P.R. China.}
 \author{H.~Lu}
 \affiliation{Key Laboratory of Particle Astrophysics, Institute 
                  of High Energy Physics, Chinese Academy of Sciences,
                  P.O. Box 918, 100049 Beijing, P.R. China.}
 \author{X.H.~Ma}
 \affiliation{Key Laboratory of Particle Astrophysics, Institute 
                  of High Energy Physics, Chinese Academy of Sciences,
                  P.O. Box 918, 100049 Beijing, P.R. China.}
 \author{G.~Mancarella}
 \affiliation{Dipartimento Matematica e Fisica "Ennio De Giorgi", 
                  Universit\`a del Salento, via per Arnesano, 73100 Lecce, Italy.}
  \affiliation{Istituto Nazionale di Fisica Nucleare, Sezione di
                  Lecce, via per Arnesano, 73100 Lecce, Italy.}
 \author{S.M.~Mari}
 \affiliation{Istituto Nazionale di Fisica Nucleare, Sezione di
                  Roma Tre, via della Vasca Navale 84, 00146 Roma, Italy.}
 \affiliation{Dipartimento di Fisica dell'Universit\`a ``Roma Tre'', 
                   via della Vasca Navale 84, 00146 Roma, Italy.}
 \author{G.~Marsella}
 \affiliation{Dipartimento Matematica e Fisica "Ennio De Giorgi", 
                  Universit\`a del Salento, via per Arnesano, 73100 Lecce, Italy.}
 \affiliation{Istituto Nazionale di Fisica Nucleare, Sezione di
                  Lecce, via per Arnesano, 73100 Lecce, Italy.}
 \author{D.~Martello}
\affiliation{Dipartimento Matematica e Fisica "Ennio De Giorgi", 
                  Universit\`a del Salento, via per Arnesano, 73100 Lecce, Italy.}
  \affiliation{Istituto Nazionale di Fisica Nucleare, Sezione di
                  Lecce, via per Arnesano, 73100 Lecce, Italy.}
 \author{S.~Mastroianni}
 \affiliation{Istituto Nazionale di Fisica Nucleare, Sezione di
                  Napoli, Complesso Universitario di Monte
                  Sant'Angelo, via Cinthia, 80126 Napoli, Italy.}
 \author{P.~Montini}
 \affiliation{Istituto Nazionale di Fisica Nucleare, Sezione di
                  Roma Tre, via della Vasca Navale 84, 00146 Roma, Italy.}
 \affiliation{Dipartimento di Fisica dell'Universit\`a ``Roma Tre'', 
                   via della Vasca Navale 84, 00146 Roma, Italy.}
 \author{C.C.~Ning}
 \affiliation{Tibet University, 850000 Lhasa, Xizang, P.R. China.}
 \author{M.~Panareo}
 \affiliation{Dipartimento Matematica e Fisica "Ennio De Giorgi", 
                  Universit\`a del Salento, via per Arnesano, 73100 Lecce, Italy.}
 \affiliation{Istituto Nazionale di Fisica Nucleare, Sezione di
                  Lecce, via per Arnesano, 73100 Lecce, Italy.}
 \author{B.~Panico}
 \affiliation{Dipartimento di Fisica dell'Universit\`a di Roma ``Tor Vergata'', via della Ricerca Scientifica 1, 00133 Roma, Italy.}
 \affiliation{Istituto Nazionale di Fisica Nucleare, Sezione di
                   Roma Tor Vergata, via della Ricerca Scientifica 1, 
                   00133 Roma, Italy.}
 \author{L.~Perrone}
 \affiliation{Dipartimento Matematica e Fisica "Ennio De Giorgi", 
                  Universit\`a del Salento, via per Arnesano, 73100 Lecce, Italy.}
 \affiliation{Istituto Nazionale di Fisica Nucleare, Sezione di
                  Lecce, via per Arnesano, 73100 Lecce, Italy.}
 \author{P.~Pistilli}
 \affiliation{Istituto Nazionale di Fisica Nucleare, Sezione di
                  Roma Tre, via della Vasca Navale 84, 00146 Roma, Italy.}
 \affiliation{Dipartimento di Fisica dell'Universit\`a ``Roma Tre'', 
                   via della Vasca Navale 84, 00146 Roma, Italy.}
 \author{F.~Ruggieri}
 \affiliation{Istituto Nazionale di Fisica Nucleare, Sezione di
                  Roma Tre, via della Vasca Navale 84, 00146 Roma, Italy.}
 \author{P.~Salvini}
 \affiliation{Istituto Nazionale di Fisica Nucleare, Sezione di Pavia, 
                  via Bassi 6, 27100 Pavia, Italy.}
 \author{R.~Santonico}
 \affiliation{Dipartimento di Fisica dell'Universit\`a di Roma ``Tor Vergata'', via della Ricerca Scientifica 1, 00133 Roma, Italy.}
 \affiliation{Istituto Nazionale di Fisica Nucleare, Sezione di
                   Roma Tor Vergata, via della Ricerca Scientifica 1, 
                   00133 Roma, Italy.}
\author{S.N. Sbano}
 \affiliation{Dipartimento Matematica e Fisica "Ennio De Giorgi", 
                  Universit\`a del Salento, via per Arnesano, 73100 Lecce, Italy.}
 \affiliation{Istituto Nazionale di Fisica Nucleare, Sezione di
                  Lecce, via per Arnesano, 73100 Lecce, Italy.}
 \author{P.R.~Shen}
 \affiliation{Key Laboratory of Particle Astrophysics, Institute 
                  of High Energy Physics, Chinese Academy of Sciences,
                  P.O. Box 918, 100049 Beijing, P.R. China.}
 \author{X.D.~Sheng}
 \affiliation{Key Laboratory of Particle Astrophysics, Institute 
                  of High Energy Physics, Chinese Academy of Sciences,
                  P.O. Box 918, 100049 Beijing, P.R. China.}
 \author{F.~Shi}
 \affiliation{Key Laboratory of Particle Astrophysics, Institute 
                  of High Energy Physics, Chinese Academy of Sciences,
                  P.O. Box 918, 100049 Beijing, P.R. China.}
\author{A.~Surdo}
 \affiliation{Istituto Nazionale di Fisica Nucleare, Sezione di
                  Lecce, via per Arnesano, 73100 Lecce, Italy.}
 \author{Y.H.~Tan}
 \affiliation{Key Laboratory of Particle Astrophysics, Institute 
                  of High Energy Physics, Chinese Academy of Sciences,
                  P.O. Box 918, 100049 Beijing, P.R. China.}
 \author{P.~Vallania}
 \affiliation{Istituto di Fisica dello Spazio Interplanetario
                   dell'Istituto Nazionale di Astrofisica, 
                   corso Fiume 4, 10133 Torino, Italy.}
 \affiliation{Istituto Nazionale di Fisica Nucleare,
                   Sezione di Torino, via P. Giuria 1, 10125 Torino, Italy.}
 \author{S.~Vernetto}
 \affiliation{Istituto di Fisica dello Spazio Interplanetario
                   dell'Istituto Nazionale di Astrofisica, 
                   corso Fiume 4, 10133 Torino, Italy.}
 \affiliation{Istituto Nazionale di Fisica Nucleare,
                   Sezione di Torino, via P. Giuria 1, 10125 Torino, Italy.}
 \author{C.~Vigorito}
 \affiliation{Istituto Nazionale di Fisica Nucleare,
                   Sezione di Torino, via P. Giuria 1, 10125 Torino, Italy.}
 \affiliation{Dipartimento di Fisica Generale dell'Universit\`a di Torino,
                   via P. Giuria 1, 10125 Torino, Italy.}
 \author{H.~Wang}
 \affiliation{Key Laboratory of Particle Astrophysics, Institute 
                  of High Energy Physics, Chinese Academy of Sciences,
                  P.O. Box 918, 100049 Beijing, P.R. China.}
 \author{C.Y.~Wu}
 \affiliation{Key Laboratory of Particle Astrophysics, Institute 
                  of High Energy Physics, Chinese Academy of Sciences,
                  P.O. Box 918, 100049 Beijing, P.R. China.}
 \author{H.R.~Wu}
 \affiliation{Key Laboratory of Particle Astrophysics, Institute 
                  of High Energy Physics, Chinese Academy of Sciences,
                  P.O. Box 918, 100049 Beijing, P.R. China.}
 \author{L.~Xue}
 \affiliation{Shandong University, 250100 Jinan, Shandong, P.R. China.}
 \author{Y.X.~Yan}
 \affiliation{Yunnan University, 2 North Cuihu Rd., 650091 Kunming, 
                   Yunnan, P.R. China.}
 \author{Q.Y.~Yang}
 \affiliation{Yunnan University, 2 North Cuihu Rd., 650091 Kunming, 
                   Yunnan, P.R. China.}
 \author{X.C.~Yang}
 \affiliation{Yunnan University, 2 North Cuihu Rd., 650091 Kunming, 
                   Yunnan, P.R. China.}
 \author{Z.G.~Yao}
 \affiliation{Key Laboratory of Particle Astrophysics, Institute 
                  of High Energy Physics, Chinese Academy of Sciences,
                  P.O. Box 918, 100049 Beijing, P.R. China.}
 \author{A.F.~Yuan}
 \affiliation{Tibet University, 850000 Lhasa, Xizang, P.R. China.}
 \author{M.~Zha}
 \affiliation{Key Laboratory of Particle Astrophysics, Institute 
                  of High Energy Physics, Chinese Academy of Sciences,
                  P.O. Box 918, 100049 Beijing, P.R. China.}
 \author{H.M.~Zhang}
 \affiliation{Key Laboratory of Particle Astrophysics, Institute 
                  of High Energy Physics, Chinese Academy of Sciences,
                  P.O. Box 918, 100049 Beijing, P.R. China.}
 \author{L.~Zhang}
 \affiliation{Yunnan University, 2 North Cuihu Rd., 650091 Kunming, 
                   Yunnan, P.R. China.}
 \author{X.Y.~Zhang}
 \affiliation{Shandong University, 250100 Jinan, Shandong, P.R. China.}
 \author{Y.~Zhang}
 \affiliation{Key Laboratory of Particle Astrophysics, Institute 
                  of High Energy Physics, Chinese Academy of Sciences,
                  P.O. Box 918, 100049 Beijing, P.R. China.}
 \author{Zhaxiciren}
 \affiliation{Tibet University, 850000 Lhasa, Xizang, P.R. China.}
 \author{Zhaxisangzhu}
 \affiliation{Tibet University, 850000 Lhasa, Xizang, P.R. China.}
 \author{X.X.~Zhou}
 \affiliation{Southwest Jiaotong University, 610031 Chengdu, 
                   Sichuan, P.R. China.}
 \author{F.R.~Zhu}
 \affiliation{Southwest Jiaotong University, 610031 Chengdu, 
                   Sichuan, P.R. China.}
 \author{Q.Q.~Zhu} 
 \affiliation{Key Laboratory of Particle Astrophysics, Institute 
                  of High Energy Physics, Chinese Academy of Sciences,
                  P.O. Box 918, 100049 Beijing, P.R. China.}
 \author{G.~Zizzi}
 \affiliation{Istituto Nazionale di Fisica Nucleare - CNAF, Viale 
                  Berti-Pichat 6/2, 40127 Bologna, Italy.}

\collaboration{ARGO-YBJ Collaboration}

\begin{abstract}
Measuring the anisotropy of the arrival direction distribution of cosmic rays provides important information on the propagation mechanisms and the identification of their sources. In fact, the flux of cosmic rays is thought to be dependent on the arrival direction only due to the presence of nearby cosmic ray sources or particular magnetic-field structures. Recently, the observation of unexpected excesses at TeV energy down to angular scale as narrow as $\sim10\deg$ raised the possibility that the problem of the origin of galactic cosmic rays may be addressed by studying the anisotropy.

The ARGO-YBJ experiment is a full-coverage EAS array, sensitive to cosmic rays with energy threshold of a few hundred GeV.
Searching for small-size deviations from the isotropy, the ARGO-YBJ collaboration explored the declination region $\delta\sim-20^{\circ}\div 80^{\circ}$, making use of about 3.7$\cdot$10$^{11}$ events collected from November 2007 to May 2012.

In this paper the detection of different significant (up to 13 standard deviations) medium-scale anisotropy regions in the arrival directions of CRs is reported. The observation was performed with unprecedented detail. 
The relative excess intensity with respect to the isotropic flux extends up to 10$^{-3}$. The maximum excess occurs for proton energies of 10-20 TeV, suggesting the presence of unknown features of the magnetic fields the charged cosmic rays propagate through, or some contribution of nearby sources never considered so far.

The observation of new weaker few-degree excesses throughout the sky region $195^{\circ}\leq$ \ra $\leq 290^{\circ}$ is reported for the first time.
\end{abstract}

\pacs{96.50.S-;95.85.Ry;96.50.sd;96.50.Bh}
\maketitle



\section{Introduction}

The measurement of anisotropy in the arrival direction of cosmic rays (CRs) is complementary to the study of their energy spectrum and chemical composition to understand their origin and propagation. It is also a tool to probe the structure of the magnetic fields through which CRs travel (for a review see, e.g. \cite{anis-rev}).

As cosmic rays are mostly charged nuclei, their trajectories are deflected by the action of the galactic magnetic field (GMF) they propagate through before reaching the Earth atmosphere, so that their detection provides directional information only up to distances as large as their gyro-radius. If CRs below $10^{15}{\rm\,eV}$ are considered and the local galactic magnetic field ($\sim3{\rm\,\mu G}$ \cite{beck01}) is accounted for, the gyro-radii are so short ($\leq 1\textrm{\ pc}$) that isotropy is expected, as no structures of the GMF are known to focus CRs within such horizon.
At most, a weak dipolar distribution may exist, reflecting the contribution of the closest CR sources.

However, a number of experiments observed an energy-dependent non-dipolar ``Large'' Scale Anisotropy (LSA) in the sidereal time frame with an amplitude of about 10$^{-4}$ - 10$^{-3}$, revealing the existence of two distinct broad regions: an excess distributed around 40$^{\circ}$ to 90$^{\circ}$ in right ascension (commonly referred to as ``tail-in'' excess, because of the position consistent with the direction of the helio-tail) and a deficit (the ``loss cone'') around 150$^{\circ}$ to 240$^{\circ}$ in right ascension (\ra) \cite{nagashima,kam07,tibet06,milagro09,eastop09,argo11,icecube11}.

The origin of this anisotropy of galactic CRs is still unknown. 
Unlike predictions from diffusion models, taking into account the role of the few closest and most recent sources (see, as an example, \cite{blasi11,erlykin06,giacinti11}), the CR arrival direction distribution in sidereal time was never found to be purely dipolar. Even two harmonics were necessary to properly describe the R.A. profiles, showing that the CR intensity has quite a complicated structure unaccountable simply by kinetic models.

Other studies suggest that a non-di-polar anisotropy could be due to a combined effect of the regular and turbulent GMF \cite{battaner09}, or to local uni- and bi-dimensional CR flows \cite{amenomori10}. The authors suggest that the LSA is generated in the interaction of galactic CRs and magnetic field in the local insterstellar space surrounding the heliosphere. 

The EAS-TOP \cite{eastop09} and IceCube \cite{icecube11} experiments observed significant anisotropy around 400 TeV. At this energy, the signal looks quite different from the modulation observed up to $\sim50$~TeV, both in amplitude (higher than expected from the extrapolation of the lower energy trend) and phase ($\sim 12\textrm{\ hrs}$ shifted); it suggests that the global anisotropy may be the superposition of different contributions from phenomena at different distances from the Earth \cite{amenomori10,desiati10}. 
On this line of thought, an underlying anisotropy related to the CR sources distribution likely manifests itself above 100 TeV, whereas at lower energy the distribution is dominated by structures at the boundaries of or contained in the solar system, as the proton gyro-radius is of the order of the helio-tail distance \cite{pogorelov09a,pogorelov09b}.
In some sense, the hundred-TeV energy range may be the one where a transition in the cause of the anisotropy occurs.

In 2007, modeling the LSA of 5 TeV CRs, the Tibet-AS$\gamma$ collaboration ran into a ``skewed'' feature over-imposed to the broad structure of the so-called tail-in region \cite{amenomori07,amenomori09}. They modeled it with a couple of intensity excesses in the plane containing the direction of the Local Insterstellar Medium velocity and the direction that neutral hydrogen enters the inner heliosheath (hydrogen deflection plane \cite{lallement05}), each of them 10$^{\circ}$-30$^{\circ}$ wide. A residual excess remained in coincidence with the helio-tail.
The same authors proposed that such a ``Medium'' Scale Anisotropy (MSA) is caused by a modulation of galactic CRs in the helio-tail \cite{amenomori11}, i.e. the region opposite to the motion of the Sun.

Afterwards the Milagro collaboration claimed the discovery of two localized regions of excess for 10 TeV CRs on angular scales of 10$^{\circ}$ with more than 12 $\sigma$ significance \cite{milagro2008}.
Regions ``A'' and ``B'', as they were named, are positionally consistent with the ``skewed feature'' observed by Tibet-AS$\gamma$ and were parametrized in terms of \ra and \dec as follows:
\begin{eqnarray*}
  {\rm region ``A":}&66^{\circ}\leq{\rm \ra}\leq 76^{\circ}&10^{\circ}\leq{\rm \dec }\leq 20^{\circ}\\
  {\rm region ``B":}&117^{\circ}\leq{\rm \ra}\leq 131^{\circ}&15^{\circ}\leq{\rm \dec }\leq 40^{\circ}\\
  &131^{\circ}\leq{\rm \ra}\leq 141^{\circ}&40^{\circ}\leq{\rm \dec }\leq 50^{\circ}.
\end{eqnarray*} 
The most intense and localized of them (as extended as 10$^{\circ}$) coincides with the direction of the helio-tail.
The average fractional excess of region ``A'' is $\sim 6\times$10$^{-4}$, whereas for region ``B'' it is $\sim$ 4$\times$ 10$^{-4}$. The Milagro collaboration excluded the hypothesis of a gamma-ray induced effect.
In both regions the spectrum is harder than the one of the isotropic part of CRs. 

The more beamed the anisotropies and the lower their rigidity, the more difficult it is to fit the standard model of CRs and galactic magnetic field to the experimental results. In this sense, the observation was rather surprising and some confirmation from other experiments was expected. In fact, the detection of a small-scale signal nested into a larger-scale modulation relies on the capability of suppressing the global CR anisotropy efficiently controlling biases in the analysis at smaller scales.

Recently, the IceCube collaboration found features compatible with the MSA also in the Southern hemisphere \cite{icecube11}. 
It is worth noting that the IceCube experiment detects muons, making us confident that charged CRs of energy above 10 TeV are observed.

To subtract the LSA superposed to the medium-scale structures, the Milagro and Icecube collaborations estimated the CR background with methods based on time-average. 
They rely on the assumption that the local distribution of the incoming CRs is slowly varying and the time-averaged signal can be used as a good estimation of the background content. 
Time-averaging methods act effectively as a high-pass filter, not allowing to study structures wider than the angle over which the background is computed in a time interval $\Delta T$ (i.e., 15$^{\circ}$/hour$\times \Delta T$ in \ra) \cite{TimeAverageBias}.

So far, no theory of CRs in the Galaxy exists yet to explain few-degree anisotropies in the rigidity region 1-10 TV leaving the standard model of CRs and that of the local magnetic field \cite{ferriere01,widrow02,page07} unchanged at the same time. 

Preliminary interpretations were based on the observation that the excesses are inside the tail-in region, and this induced the authors to consider interactions of CRs with the heliosphere \cite{milagro2008}.
Several authors, noticing that the TeV region is usually free of heliosphere-induced effects, proposed a model where the excesses are produced in the Geminga supernova explosion \cite{SalvatiSacco}. In the first variant of the model, CRs simply diffuse (Bohm regime) up to the solar system, whereas the second version limits the diffusion to the very first phase of the process and appeals to non-standard divergent magnetic field structure to bring them to the Earth. 
Other people \cite{DruryAharonian} proposed similar schemes involving local sources and magnetic traps guiding CRs to the Earth. It must be noticed that sources are always intended not to be farther than 100 - 200 pc. Moreover the position of the excesses in galactic coordinates, symmetrical with respect to the galactic plane, played an important role in inspiring such models.

Based on the observation that all nearby sources and new magnetic structures brought in to explain the medium-scale anisotropy should imply other experimental signatures that had been observed in the past, some other  models were proposed. 
In \cite{amenomori10,amenomori11} the authors suggest that the magnetic field in the helio-tail (that is, within $\sim$70 AU to $\sim$340 AU from the Sun) is responsible for the observed midscale anisotropy in the energy range 1 - 30 TeV. In particular, this effect is expressed as two intensity enhancements placed along the hydrogen deflection plane, each symmetrically centered away from the helio-tail direction.

The hypothesis that the effect could be related to the interaction of isotropic CRs with the heliosphere has been re-proposed in \cite{desiati10}. Grounding on the coincidence of the most significant localized regions with the helio-spheric tail, magnetic reconnection in the magneto-tail has been shown to account for beaming particles up to TeV energies. 
On a similar line, in \cite{desiati11} the authors suggest that even the LSA below 100 TeV is mostly due to particle interactions with the turbulent ripples generated by the interaction of the helio-spheric and interstellar magnetic field. This interaction could be the dominant factor that re-distributes the LSA from an underlying existent anisotropy. At the same time, such a scattering can produce structures at small angular scale along lines of sight that are almost perpendicular to the local interstellar magnetic fields.

Recently, the role the GMF turbulent component on TeV - PeV CRs was emphasized to explain the MSA \cite{giacinti11}. The authors show that energy-dependent medium and small scale anisotropies necessarily appear, provided that there exists a LSA, for instance from the inhomogeneous source distribution. The small-scale anisotropies naturally arise from the structure of the GMF turbolence, typically within the CR scattering length.

It was also suggested that CRs might be scattered by strongly anisotropic Alfven waves originating from turbulence across the local field direction \cite{MalkovEtAl}.

Besides all these ``ad hoc'' interpretations, several attempts occurred in trying to insert the CR excesses in the framework of recent discoveries from satellite-borne experiments, mostly as far as leptons are concerned. In principle there is no objection in stating that few-degree CR anisotropies are related to the positron excess observed by Pamela, Fermi and AMS \cite{PamelaPositrons,FermiPositrons,AMSPositrons,Kistler:2012ag,Gaggero:2013rya}. All observations can be looked at as different signatures of common underlying physical phenomena.

The ARGO-YBJ collaboration reports here the observation of a medium scale CR anisotropy in the 1 - 30 TeV energy region.
The evidence of new weaker few-degree excesses throughout the sky region $195^{\circ}\leq$ \ra $\leq 290^{\circ}$ is reported for the first time. The analysis presented in this paper relies on the application of time-average based method to filter out the LSA signal. The analysis of the CR anisotropy in the harmonic space, based on the standard spherical-harmonic transform or on the needlet transform \cite{needlet}, will be the subject of a forthcoming paper.

The paper is organized as follows. In the section II the detector is described and its performance summarized. In  the section III the details of the ARGO-YBJ data analysis are summarized. The results are presented and discussed in  the section IV. Finally, the conclusions are given in  the section V.

\section{The ARGO-YBJ experiment}

The ARGO-YBJ experiment, located at the YangBaJing Cosmic Ray Laboratory (Tibet, P.R. China, 4300 m a.s.l., 606 g/cm$^2$, geographic latitude $30^\circ\,06^{\prime}\,38^{\prime\prime}\ {\rm N}$), is made of a central carpet $\sim$74$\times$ 78 m$^2$, made of a single layer of Resistive Plate Chambers (RPCs) with $\sim$93$\%$ of active area, enclosed by a partially instrumented guard ring ($\sim$20$\%$) up to $\sim$100$\times$110
m$^2$. 
The RPC is a gaseous detector working with uniform electric field generated by two parallel electrode plates of high bulk resistivity (10$^{11}\Omega$ cm). The intense field of 3.6 kV/mm at a 0.6 atm pressure provides very good time resolution (1.8 ns) and the high electrode resistivity limits the area interested by the electrical discharge to few mm$^2$.
The apparatus has a modular structure, the basic data-acquisition sector being a cluster (5.7$\times$7.6 m$^2$), made of 12 RPCs (2.85$\times$1.23 m$^2$ each). Each chamber is read by 80 external strips of 6.75$\times$61.8 cm$^2$ (the spatial pixel), logically organized in 10 independent pads of 55.6$\times$61.8 cm$^2$ which represent the time pixel of the detector \cite{aielli06}. 
The read-out of 18360 pads and 146880 strips are the experimental output of the detector. 
The RPCs are operated in streamer mode by using a gas mixture (Ar 15\%, Isobutane 10\%, TetraFluoroEthane 75\%) for high altitude operation \cite{bacci00}. The high voltage settled at 7.2 kV ensures an overall efficiency of about 96\% \cite{aielli09a}.
The central carpet contains 130 clusters (hereafter ARGO-130) and the full detector is composed of 153 clusters for a total active
surface of $\sim$6700 m$^2$. The total instrumented area is $\sim$11000 m$^2$.

A simple, yet powerful, electronic logic has been implemented to build an inclusive trigger. This logic is based on a time correlation between the pad signals depending on their relative distance. In this way, all the shower events giving a number of fired pads N$_{pad}\ge$ N$_{trig}$ in the central carpet in a time window of 420 ns generate the trigger.
This trigger can work with high efficiency down to N$_{trig}$ = 20, keeping the rate of random coincidences negligible.
The time calibrations of the pads is performed according to the method reported in \cite{hhh07,aielli09b}.

The whole system, in smooth data taking since July 2006 with ARGO-130, has been in stable data taking with the full apparatus of 153 clusters from November 2007 to January 2013, with the trigger condition N$_{trig}$ = 20 and a duty cycle $\geq$85\%. The trigger rate is $\sim$3.5 kHz with a dead time of 4$\%$.

Once the coincidence of the secondary particles has been recorded,
the main parameters of the detected shower are reconstructed following the procedure described in \cite{bartoli11a}. 
In short, the reconstruction is split into the following steps. Firstly, the shower core position is derived with the Maximum Likelihood method from the lateral density distribution of the secondary particles. In the second step, given the core position, the shower axis is reconstructed by means of an iterative un-weighted planar fit able to reject the time values belonging to non-gaussian tails of the arrival time distribution. Finally, a conical correction is applied to the surviving hits in order to improve the angular resolution.
Details on the analysis procedure (e.g., reconstruction algorithms, data selection, background evaluation, systematic errors) are discussed in \cite{aielli10,bartoli11a,bartoli11b}.

The performance of the detector (angular resolution, pointing accuracy, energy scale calibration) and the operation stability are continuously monitored by observing the Moon shadow, i.e., the deficit of CRs detected in its direction \cite{bartoli11a,bartoli12a}. 
ARGO-YBJ observes the Moon shadow with a sensitivity of $\sim$9 standard deviations (s.d.) per month. 
The measured angular resolution is better than 0.5$^{\circ}$ for CR-induced showers with energy E $>$ 5 TeV and the overall absolute pointing accuracy is $\sim$0.1$^{\circ}$.
The absolute pointing of the detector is stable at a level of 0.1$^{\circ}$ and the angular resolution is stable at a level of 10\% on a monthly basis.
The absolute rigidity scale uncertainty of ARGO-YBJ is estimated to be less than 13\% in the range 1 - 30 TeV/Z \cite{bartoli11a,bartoli12a}.
The last results obtained by ARGO-YBJ are summarized in \cite{gdisciascio12}.

\section{Data analysis}
\label{sec:datanalysis}

The analysis reported in this paper used $\sim$3.70$\times$10$^{11}$ showers recorded by the \argo experiment from November $8^{\rm th}$, 2007 till May $20^{\rm th}$, 2012, after the following selections: (1) $\geq$25 strips must be fired on the ARGO-130 central carpet; (2) zenith angle of the reconstructed showers $\leq$50$^{\circ}$; (3) reconstructed core position inside a 150$\times$150 m$^2$ area centered on the detector.
Data have been recorded in 1587 days out of 1656, for a total observation time of 33012 hrs ($86.7\%$ duty-cycle). A selection of high-quality data reduced the data-set to 1571 days. 
The zenith cut selects the \dec region $\delta\sim$ -20$^{\circ}\div$ 80$^{\circ}$.
According to the simulation, the median energy of the isotropic CR proton flux is E$_p^{50}\approx$1.8 TeV (mode energy $\approx$0.7 TeV).
No gamma/hadron discrimination algorithms have been applied to the data. Therefore, in the following the sky maps are filled with all CRs possibly including photons. Several off-line methods to separate gamma from hadrons, as well as to distinguish low-mass and high-mass CRs are currently under study.

In order to investigate the energy dependence of the observed phenomena, the data-set has been divided into five multiplicity intervals. The Table \ref{tab:rates} reports the size boundaries and the amount of events for each interval.

\begin{figure}[!htbp]
\epsfxsize=9cm
  \includegraphics[angle=0,width=0.45\textwidth]{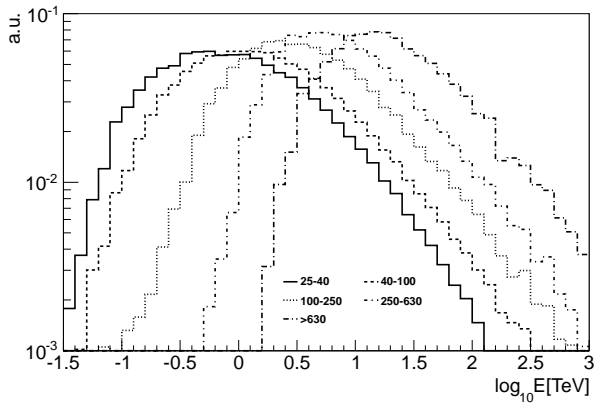}
  \caption{Energy distribution of proton-initiated showers firing as many strips as numbers reported in Table \ref{tab:rates}. Showers are simulated to be initiated by protons with energy spectrum as in \cite{Horandel}.}
  \label{fig:energy_distribution}
\end{figure}

As a reference value, the right column reports the median energy of isotropic CR protons for each multiplicity interval obtained via Monte Carlo simulation. This choice is inherited from the standard LSA analyses, but it can be only approximately interpreted as the energy of the CRs giving the MSA. In fact the elemental composition and the energy spectrum of these structures are not known and that of CR protons is just an hypothesis. The energy distribution of protons sampled according to \cite{Horandel} is also given in the Fig. \ref{fig:energy_distribution} for each multiplicity band.

The ARGO-YBJ acceptance distribution for the full data-set is essentially the same as reported in \cite{bartoli11a}. In the investigated energy range the light component (p+He) accounts for more than 90\% of the triggers, being the contribution of heavier nuclei less than 10\% (see also \cite{bartoli12b}). The photon effective areas are published in \cite{aielli09c}.

In addition, as it will be discussed in more detail, the multiplicity-energy relation is a function of the declination, which is difficult to be accounted for for sources as extended as 20$^{\circ}$ or more. 
%
\begin{table}
  \centering
  \begin{tabular}{l|cc|c}
    {\bf Strip-multiplicity}&{\bf number of}& &{$\mathbf{E_p^{50}{\rm\ [TeV]}}$}\\
    {\bf interval}&{\bf events}&&\\
    \hline
    $25-40$&$1.41673\times10^{11}$&$(38\%)$&0.66\\
    $40-100$&$1.75695\times10^{11}$&$(48\%)$&1.4\\
    $100-250$&$3.80812\times10^{10}$&$(10\%)$&3.5\\
    $250-630$&$1.09382\times10^{10}$&$(3\%)$&7.3\\
    more than $630$&$4.34442\times10^{9}$&$(1\%)$&20
  \end{tabular}
  \caption[Strip-multiplicity intervals used in the analysis.]{Multiplicity intervals used in the analysis. The central columns report the number of collected events. The right column shows the corresponding isotropic CR proton median energy.}
  \label{tab:rates}
\end{table}
%

The background contribution has been estimated with the Direct Integration and the Time Swapping methods \cite{Fleysher, bartoli11a}. 
In both cases, to minimize the systematic effects due to the environmental parameters variations \cite{aielli10,bartoli13}, the integration time has been set to $\Delta T=3$~hrs. For the Direct Integration method the size of the sky pixel is $0.07\deg\times0.07\deg$ and the time bin is $12$~sec wide. As far as the Time Swapping technique is concerned, the oversampling factor has been set to 10. We recall that no structures wider than $45\deg$ in \ra can be inspected as $\Delta T=3$~hrs. These two methods are equivalent each other. The analysis of the anisotropies has been carried out applying both techniques and no differences greater than 0.3 standard deviations have been observed. Because of its higher oversampling factor (i.e. smaller background fluctuations), all the results reported in the following are obtained with the Direct Integration method. Detailed studies of the efficiency of time-average based methods in filtering out large-scale structures in the CR arrival distribution were performed recently in \cite{TimeAverageBias}. Potential biases on the intensity, the position and the morphology induced by these techniques have been investigated. The result is that the large-scale anisotropy, as known from literature, is well filtered out (residual effect less than $10\%$); moreover, the intensity of the medium scale signal is attenuated at most by a factor of $15\%$ and the effect on the CR arrival direction (i.e. on the position and the morphology) is negligible if compared with the angular resolution. All these values are to be considered as upper limits, because the result of the estimation depends on the intensity of the \lsa in the sky region under consideration. For these reasons, the systematic uncertainty on the intensity estimation induced by the analysis techniques is about $20\%$.

To proceed with a blind search of extended CR features, no variation of the angular scale was performed to maximize the significance.
The sky maps have been smoothed by using the detector Point Spread Function (PSF). For each multiplicity interval, the PSF has been computed with Monte Carlo simulations, where the CR composition given in \cite{Horandel} was used to build the signal. The dependence of the PSF on the arrival direction was not considered, because the simulations show strong stability up to $\theta=45\deg$ and reproduce the angular resolution quite well in the range $\theta=45\deg$-50$\deg$ (the zenith range $\theta=45\deg-50\deg$ contains less than 8\% of the events). In this respect, the PSF averaged over the zenith angle range $\theta<50\deg$ was found not to introduce systematics. A validation of the reconstruction algorithms and of the Monte Carlo predictions about the angular resolution came from the study of the Moon shadow, reported in \cite{bartoli11a}. 

The PSF was not optimized for a particular chemical species, neither for any \emph{ad-hoc} angular scale. As there are no hypotheses \emph{a priori} on the nature or the extension or even the energy spectrum of the phenomenon, the analysis can be defined as a ``blind search'' of excesses and deficits with respect to the estimated background. The significance of the detection was evaluated in the following way. The estimated background map $b$ was used as seed to generate 10$^{12}$ random ``no-signal'' maps $e$, according to the Poissonian distribution. For each of them, the PSF smoothing was applied and for every pixel the residual $N_e-N_b$ in units of $\sigma=\sqrt{\sigma_e^2+\sigma_b^2}$ was computed. Each residual from every pixel from all the random maps was collected in a distribution, which was found to be a gaussian with mean $\mu=(1.1\pm1.3)10^{-7}$ and rms $\sigma=1.09432\pm0.00026$. The difference of the last value from 1 is due to the correlation between pixels induced by smoothing, as the same procedure applied to non-smoothed data gave $1-\sigma= (-5\pm 6)\cdot 10^{-5}$. The distribution obtained with this method was used to compute chance p-values for each experimental map. These p-values were traduced in s.d. units to represent the maps as shown in the next section.
%
%
\section{Results}
\begin{figure*}[!htbp]
\epsfxsize=9cm  \includegraphics[angle=0,width=0.8\textwidth]{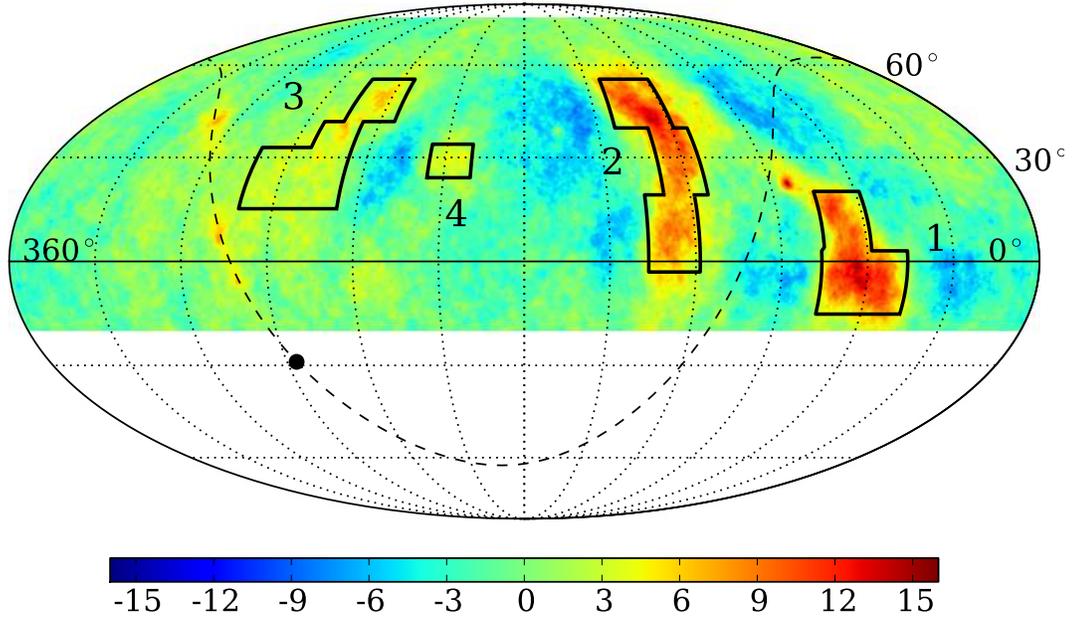}\\
(a)\\
\includegraphics[angle=0,width=0.8\textwidth]{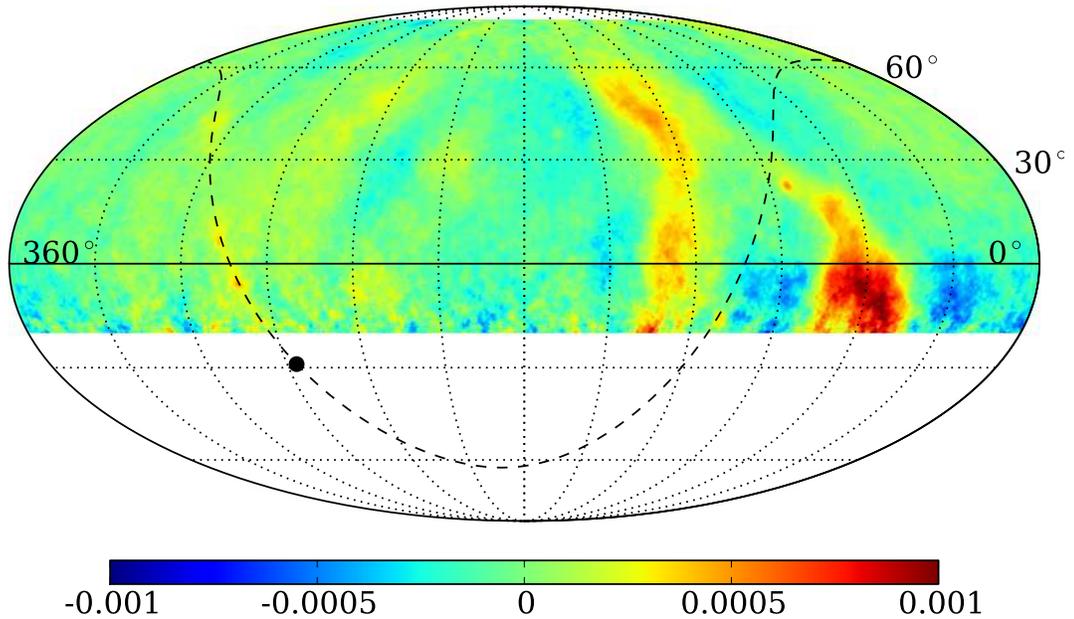}\\
(b)\\
  \caption{ARGO-YBJ sky-map in equatorial coordinates for events with N$_{strip}>$ 25. The maps have been smoothed with an angle given by the PSF of the detector. {\em Plot (a):} statistical significance of the observation in s.d.. The boxes represent the parametrization of the regions of interest (see the section \ref{sec:localization} and the Table \ref{tab:region_parametrization}). {\em Plot (b):} relative excess with respect to the estimated background. The dashed line represents the Galactic Plane and the black point the Galactic Center. \emph{Mollweide projection} on the Healpix pixelization scheme \cite{healpix}.}
  \label{fig:figA}
\end{figure*}

%
The Fig. \ref{fig:figA} shows the ARGO-YBJ sky map in equatorial coordinates as obtained with all selected events. 
The upper plot shows the statistical significance of the observation while the lower one shows the relative excess with respect to the estimated background.
They look slightly different because of the atmosphere thickness that the showers must cross before triggering the apparatus, increasing with the arrival zenith angle. As a consequence most significant regions do not necessarily coincide with most intense excesses. It should be noticed that also gamma-ray-induced signals are visible, because no gamma/hadron separation is applied.

The most evident features are observed by ARGO-YBJ around the positions $\alpha\sim$ 120$^{\circ}$, $\delta\sim$ 40$^{\circ}$ and $\alpha\sim$ 60$^{\circ}$, $\delta\sim$ -5$^{\circ}$, spatially consistent with the regions detected by Milagro \cite{milagro2008}. These regions are observed with a statistical significance of about 15 s.d. and are represented on the significance map together with the other regions of interest described in this paper (see the section \ref{sec:localization} and the Table \ref{tab:region_parametrization}).
As known from literature \cite{milagro2008, TimeAverageBias}, the deficit regions parallel to the excesses are due to using also the excess events to evaluate the background, which turns out to be overestimated. Symmetrically, deficit regions, if any, would be expected to be surrounded by weaker excess halos, which were not observed.
On the left side of the sky map, several new extended features are visible, though less intense than the ones aforementioned. The area $195^{\circ}\leq \ra \leq 290^{\circ}$ seems to be full of few-degree excesses not compatible with random fluctuations (the statistical significance is up to 7 s.d.). 
The observation of these structures is reported here for the first time.
%
%

\begin{figure}[!htbp]
\epsfxsize=9cm
  \includegraphics[angle=0,width=0.45\textwidth]{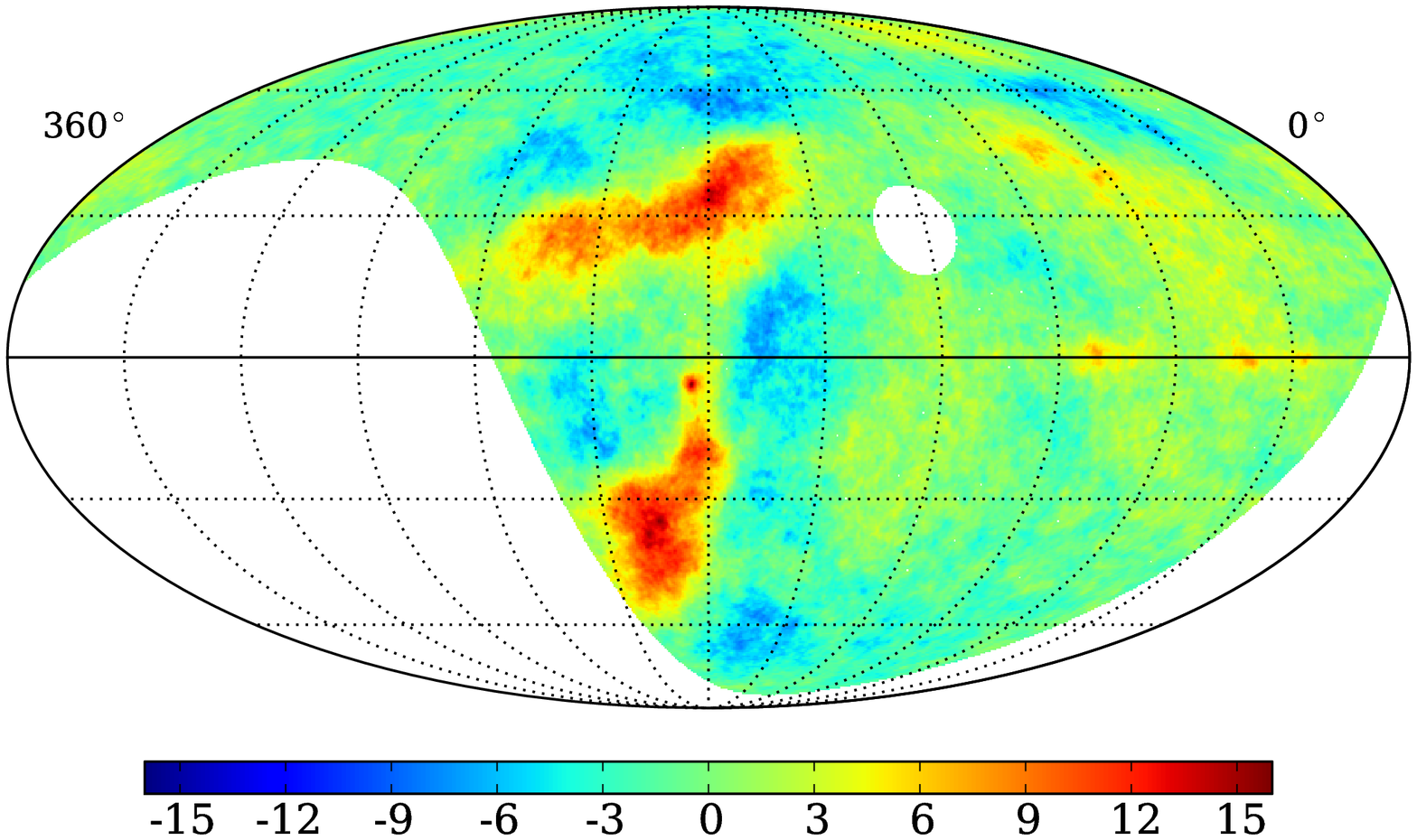}
  \caption{ARGO-YBJ sky-map of Fig. \ref{fig:figA}(a) in galactic coordinates. The map center points towards the anti-center.}
  \label{figD}
\end{figure}

The upper plot of Fig. \ref{fig:figA} is represented in galactic coordinates in Fig. \ref{figD}. As it is clearly visible in this figure, the hot spots 1 and 2 are distributed symmetrically with respect to the Galactic plane and have longitude centered around the galactic anti-center. 
The new detected hot spots do not lie on the galactic plane and one of them is very close to the galactic north pole.

%
%
\subsection{Localization of the MSA regions}
\label{sec:localization}
Looking at the map of Fig. \ref{fig:figA}, apart from the Galactic Plane, where the gamma-ray sources Crab Nebula, MGRO J1908+06 \cite{argo1908}, MGRO J2031+41 \cite{argo2031}, HESS J1841-055 \cite{argo1841} are  visible, four regions have a significance greater than 5 s.d.
%
\begin{figure*}[!htbp]
\centering
\subfigure[]{\includegraphics[width=0.45\textwidth]{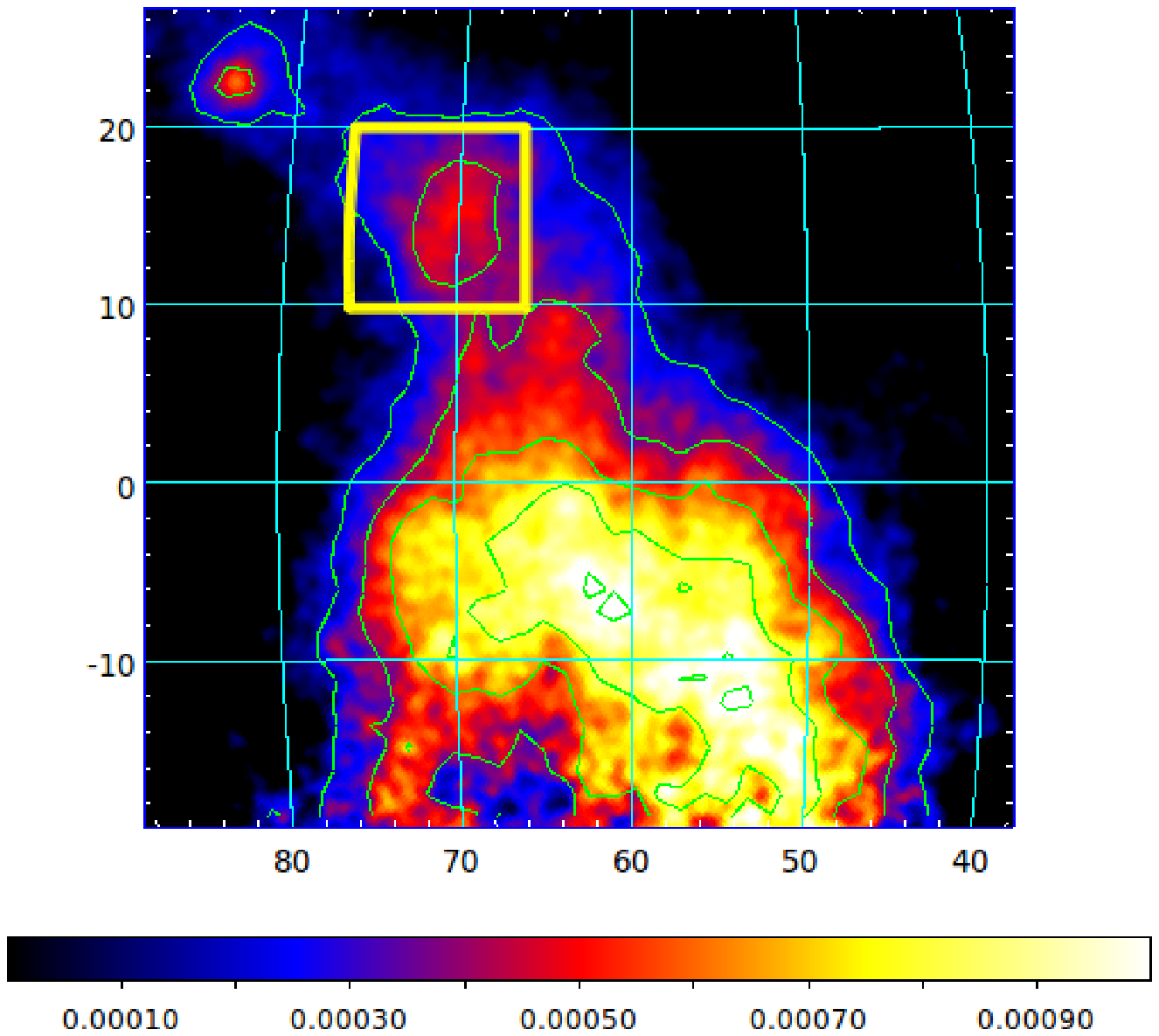}\label{fig:msa-closer-1}}
\subfigure[]{\includegraphics[width=0.35\textwidth]{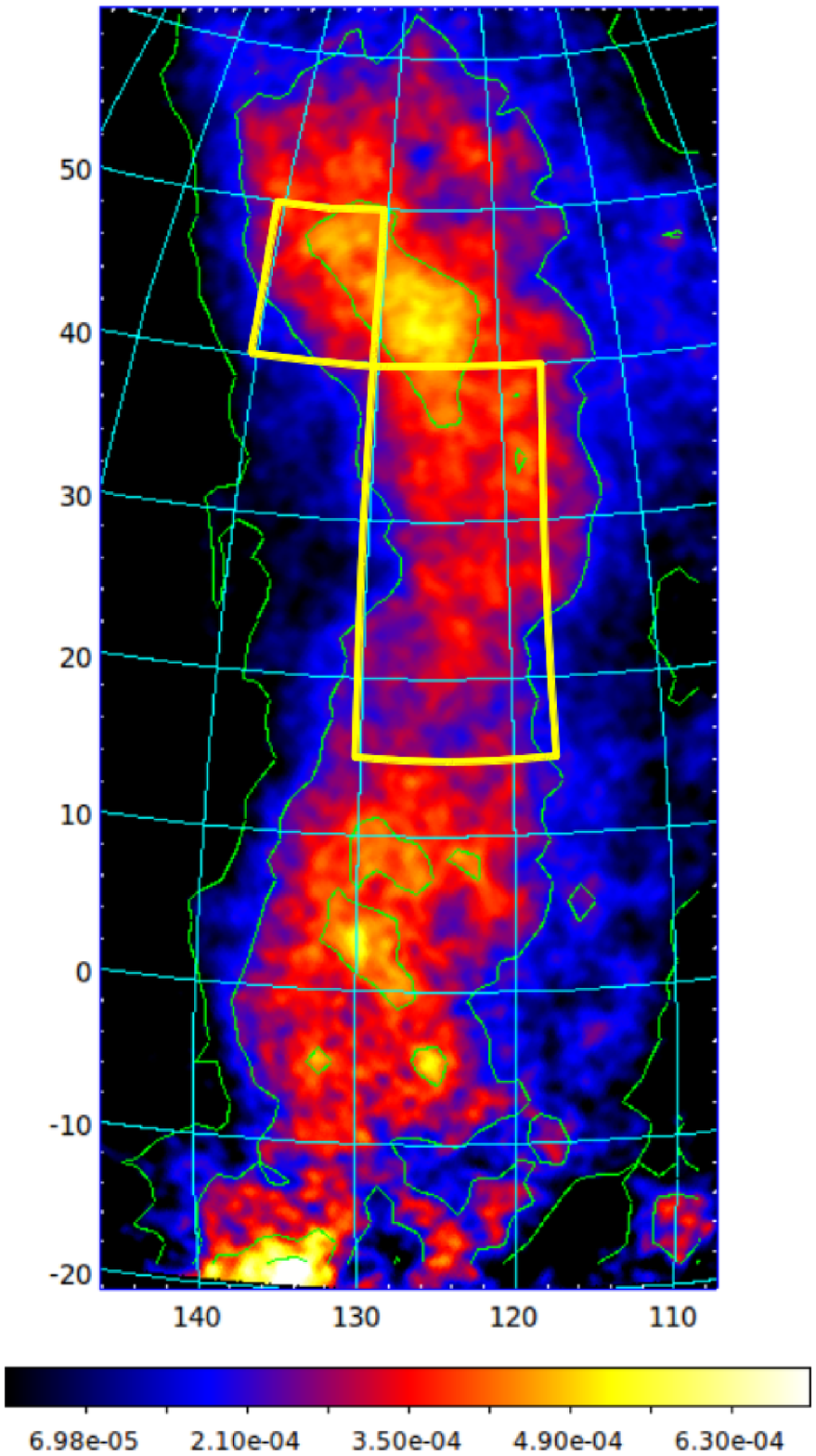}\label{fig:msa-closer-2}}
\subfigure[]{\includegraphics[width=0.45\textwidth]{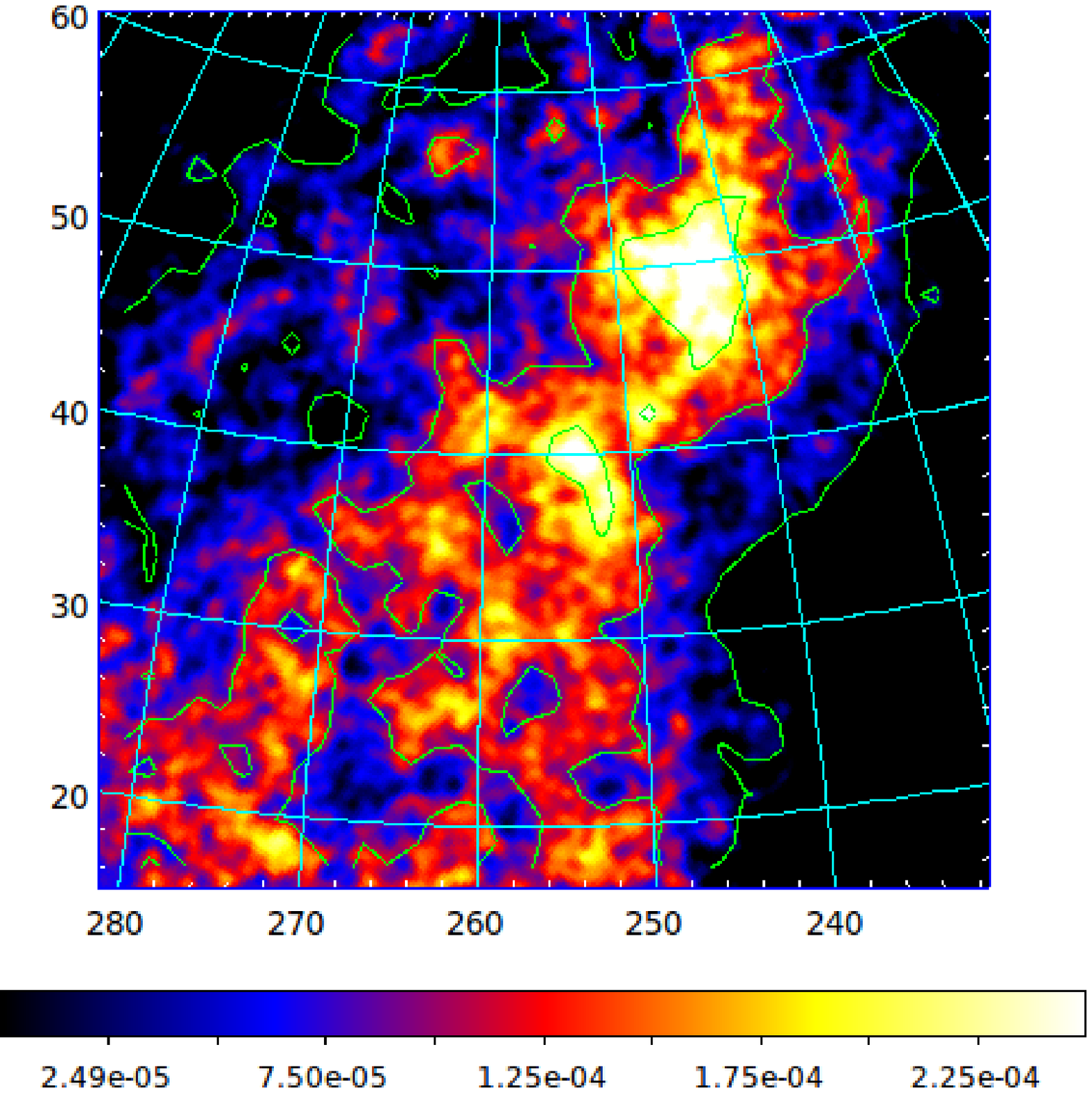}\label{fig:msa-closer-3}}
\subfigure[]{\includegraphics[width=0.45\textwidth]{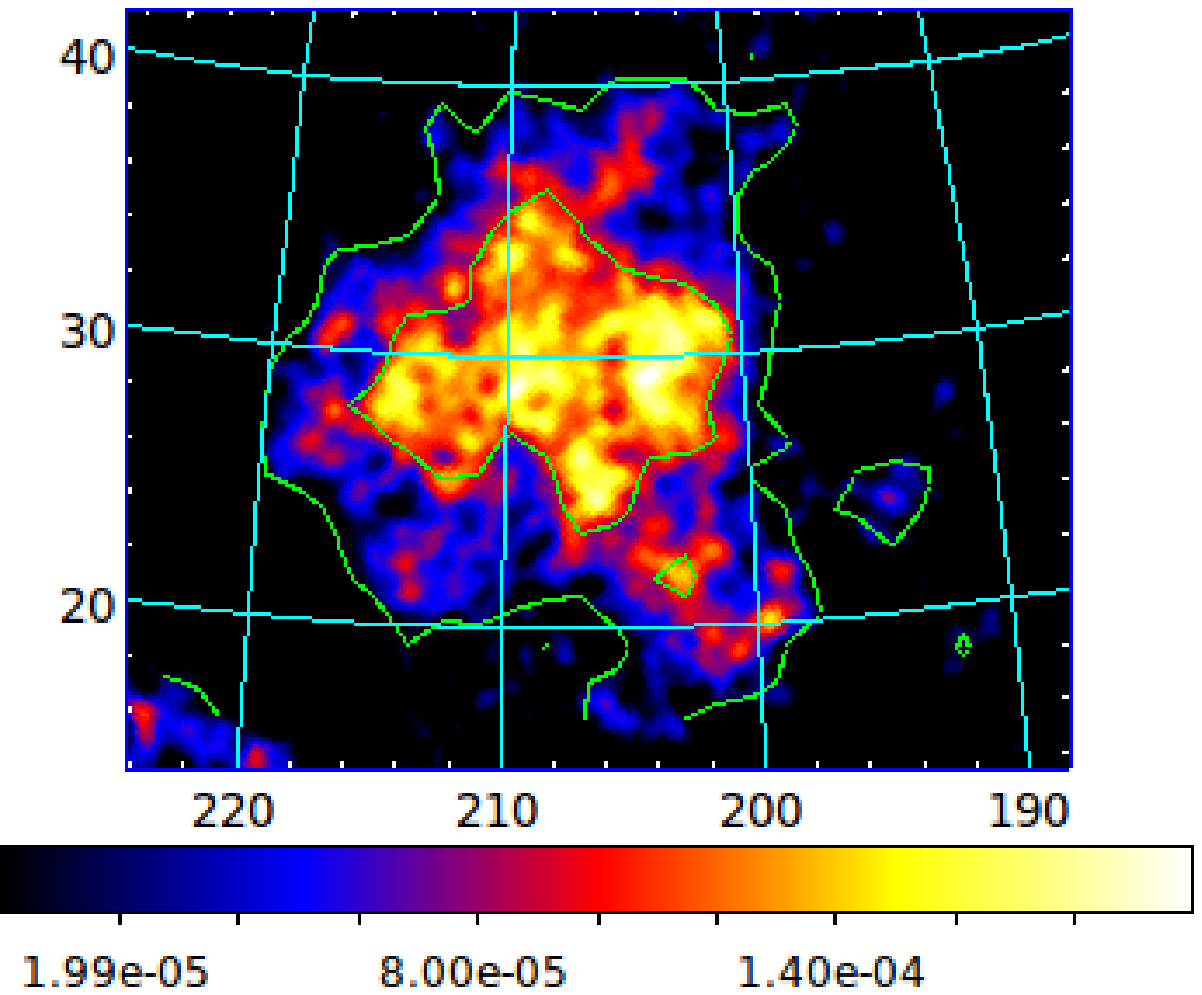}\label{fig:msa-closer-4}}
\caption{Different \msa regions observed by ARGO-YBJ. The relative excess with respect to the estimated background is shown. In the figures \ref{fig:msa-closer-1} and \ref{fig:msa-closer-2} the Milagro's regions ``A'' and ``B'' are represented with the $\ra-\dec$ boxes. The regions \ref{fig:msa-closer-3} and \ref{fig:msa-closer-4} are observed for the first time by ARGO-YBJ with a statistical significance greater than 5 s.d. . Data are represented in equatorial coordinates. Contour lines are drawn for excess intensity $(0.0,0.2,0.4,0.6,0.8)\,10^{-3}$ for regions 1 and 2, $(0.0,0.1,0.2)\,10^{-3}$ for regions 3 and 4. The yellow boxes correspond to the shapes given by the Milagro experiment for regions ``A'' and ``B'' \cite{milagro2008}. The localized Crab Nebula excess is visible in the upper-left part of the \ref{fig:msa-closer-1} (see text).}
\label{fig:msa-closer}
\end{figure*}
%

The Fig. \ref{fig:msa-closer-1} represents the zoom of the most intense visible excess (``region 1'', hereafter). The detection significance is greater than 16 s.d. and the excess intensity reaches $10^{-3}$. The Crab Nebula excess ($(\ra,\dec)=(83.66^{\circ},22.04^{\circ})$) is visible in the upper-left part of the figure. The intensity of the signal is in agreement with the expectation from simulation, once the details of the analysis are accounted for, mostly the contribution of the source (which is not excluded) and the PSF, optimized for CRs instead of $\gamma$~-rays.

The Fig. \ref{fig:msa-closer-2} represents the zoom of the most extended excess (``region 2'', hereafter). The detection significance is about 15 s.d. and the excess intensity reaches $5.0\ 10^{-4}$.

The Fig. \ref{fig:msa-closer-3} represents the zoom of a wide excess, less significant than the two regions already seen (``region 3'', hereafter). It has quite composite a morphology, with the most intense part of the signal within $10-15\deg$ around $(\ra,\dec)=(240\deg,45\deg)$. The detection significance is about 7 s.d. and the excess intensity reaches $2.3\ 10^{-4}$. This region contains the Mrk501 gamma-ray source \cite{argoMRK501}, whose contribution is not considered in this work. All the events coming from a direction whitin $3\deg$ around the source nominal position ($\ra = 253.47\deg$ and $\dec=39.76\deg$) are rejected.

The Fig. \ref{fig:msa-closer-4} represents the zoom of a small excess, the least significant observed by \argo (``region 4'', hereafter). The detection significance is 5.5 s.d. and the excess intensity reaches $1.6\ 10^{-4}$.

While regions 1 and 2 can be looked at as the excesses reported in the previous literature (the ``skewed'' feature by Tibet-AS$\gamma$ as well as the region ``A'' and ``B'' by Milagro \cite{tibet06,milagro2008}), 
regions 3 and 4 are observed here for the first time. 

Actually, even about regions 1 and 2 the \argo data analysis uncovers unexpected morphological and energetic aspects. In addition it shows details unaccessible to any previous experiment. 
In fact, in order to determine the energy spectrum, the Milagro collaboration parameterized the source morphology with three different rectangles in the (\ra,\dec) space.
They are represented in Fig. \ref{fig:msa-closer}, superimposed on the \argo intensity map. What is most noticeable therein, is that the adopted parameterization hardly fits the excesses observed by \argo.

As far as the region ``B'' is concerned, although defined in a composite way, with two rectangles to follow its shape, the most intense part of the signal lays out of the yellow boxes. The same holds for the lowest \dec zone of the excess. Moreover, the rectangles seem to be narrower in \ra than the angular span observed by \argo. It may be due to the different time interval over which the background is computed. In fact, the Milagro collaboration used 2 hrs, being sensitive only to features less than $30\deg$ wide in \ra.

Coming to the region ``A'' (Fig. \ref{fig:msa-closer-1}) we see that only the highest \dec part of the region is enclosed in the yellow box (the one closest to the Crab Nebula). The rest of the signal is missed in the parameterization of Milagro. That is  likely due to the applied smoothing technique by the experimenters. In fact, Milagro adopts a $10\deg$ top-hat as smoothing kernel, which means that all pixels closer than $10\deg$ to the \dec boundary have to be rejected in the analysis. Since the Milagro site latitude was $\sim35.9\deg$ and they analyzed events arrived within $45\deg$ zenith angle, the minimum reachable \dec was $\sim 0\deg$. This value is just the upper boundary of the most intense part of the region 1 (see the Fig. \ref{fig:msa-closer-1}).

Some features of the \argo experiment, as well as the choices adopted in this analysis, allow to go past the limits of the previous observations. In fact, as already said, $\Delta T=3{\rm\ hrs}$ has been used in this analysis, what allows to be sensitive up to $45\deg$ in \ra. Moreover, the \argo site latitude is $\sim30\deg$, the zenith angle selection cut is $\theta<50\deg\ $  and the PSF is used as the smoothing kernel, all that allowing to push the limits of the observable sky down to  \dec values as low as $\delta\sim-20\deg$.

As a consequence, to determine the energy of the four detected excesses, a suitable parameterization of their morphology has been introduced. For the sake of simplicity, all regions have been modeled with ``boxes'' in the (\ra,\dec) space. In case of complex shapes, a composition of boxes is used. The boxes are represented in the Fig. \ref{fig:figA}~(a) and their boundaries are reported in Table \ref{tab:region_parametrization}. They select the part of signal more than 3 s.d..
%
\begin{table*}[!htpb]
\centering
\begin{tabular}{l||c|c||c|c||r}
{\bf Region}&{\bf Lowest}&{\bf Highest}&{\bf Lowest}&{\bf Highest}&{\bf Sub-region}\\
{\bf name}&{\bf \ra}&{\bf \ra}&{\bf \dec}&{\bf \dec}&{\bf name}\\
\hline
\hline
\multirow{2}{*}{Region 1}&$58.5\deg$&$75.5\deg$&$3\deg$&$20\deg$&Region 1U\\
&$46\deg$&$76\deg$&$-15\deg$&$3\deg$&Region 1L\\
\hline
\multirow{3}{*}{Region 2}&$119\deg$&$143\deg$&$39\deg$&$55\deg$&Region 2U\\
&$113.5\deg$&$129.5\deg$&$19\deg$&$39\deg$&Region 2M\\
&$118.5\deg$&$136.5\deg$&$-3\deg$&$19\deg$&Region 2L\\
\hline
\multirow{3}{*}{Region 3}&$234\deg$&$255\deg$&$41\deg$&$55\deg$&Region 3U\\
&$247\deg$&$263\deg$&$33\deg$&$41\deg$&Region 3M\\
&$247\deg$&$282\deg$&$15\deg$&$33\deg$&Region 3L\\
\hline
Region 4&$200\deg$&$216\deg$&$24\deg$&$34\deg$&
\end{tabular}
\caption{Parameterization of the four MSA regions.}
\label{tab:region_parametrization}
\end{table*}
%
Apart from the region 4, which looks quite uniform, all the observed anisotropies appear to be made of some sub-regions.

In the significance map of the region 1, two zones can be distinguished: they correspond to the regions 1U and 1L of the Table \ref{tab:region_parametrization}. The region 1U hosts the signal detected by Milagro in their region ``A'', although it is slightly larger. Actually, if the intensity map is considered as a whole, the excess is found to be uniformly distributed, so that the subregions do not seem so distinct.

Concerning the region 2, its morphology is so complex that three boxes were needed to enclose it. Both in the significance and the ratio map, the excess appears to be made of two distinct hot spots, above and below $\delta\sim15\deg$. 

The region 3 is quite a difficult zone, as it is made of an intense excess in the highest \dec part, that is divided in turn in two hot spots. They are modeled as region 3U and region 3M respectively. The most extended part of the emission is located in the lowest \dec side of the region, less intense and quite uniform. We named it region 3L. 

Table \ref{tab:region_event_number} reports the number of events collected inside each anisotropy region, together with the extension (solid angle) of the assumed modelization.

We note that the regions over which \argo observes significant medium scale anisotropies have total extension $\sim0.8$~sr, i.e one third of the \argo field of view in celestial coordinates. 
%
\begin{table*}[!htpb]
\centering
\begin{tabular}{l||c|c||c}
{\bf Region}&{\bf Collected}&{\bf Background}&{\bf Region}\\
{\bf name}&{\bf events $(\times10^9)$}&{\bf est. events $(\times10^9)$}&{\bf extension (sr)}\\
\hline
\hline
{\bf Region 1}&7.49476&7.49139&0.2489\\
Region 1U&4.39145&4.39001&0.0860\\
Region 1L&3.10209&3.10015&0.1629\\
\hline
{\bf Region 2}&10.5233&10.5199& 0.2550\\
Region 2U&5.21602&5.21415& 0.0795\\
Region 2M&6.18423&6.18236& 0.1083\\
Region 2L&5.30654&5.30495& 0.0848\\
\hline
{\bf Region 3}&17.8248&17.8226& 0.2639*\\
Region 3U&2.68946&2.68902& 0.0598\\
Region 3M&1.20453&1.20442& 0.0311\\
Region 3L&11.8031&11.8019& 0.1816\\
\hline
{\bf Region 4}&3.03260&3.03224&0.0426
\end{tabular}
\caption[Events collected for each anisotropy region.]{Events collected for each anisotropy region. The solid angle used to model the regions is reported too. The extension of the region 3 is after subtracting Mrk501 (3U+3M+3L=0.2725 sr).}
\label{tab:region_event_number}
\end{table*}
%

\subsection{The multiplicity spectrum}

\begin{figure}[!htbp]
\centering
{\includegraphics[angle=0,width=.4\textwidth]{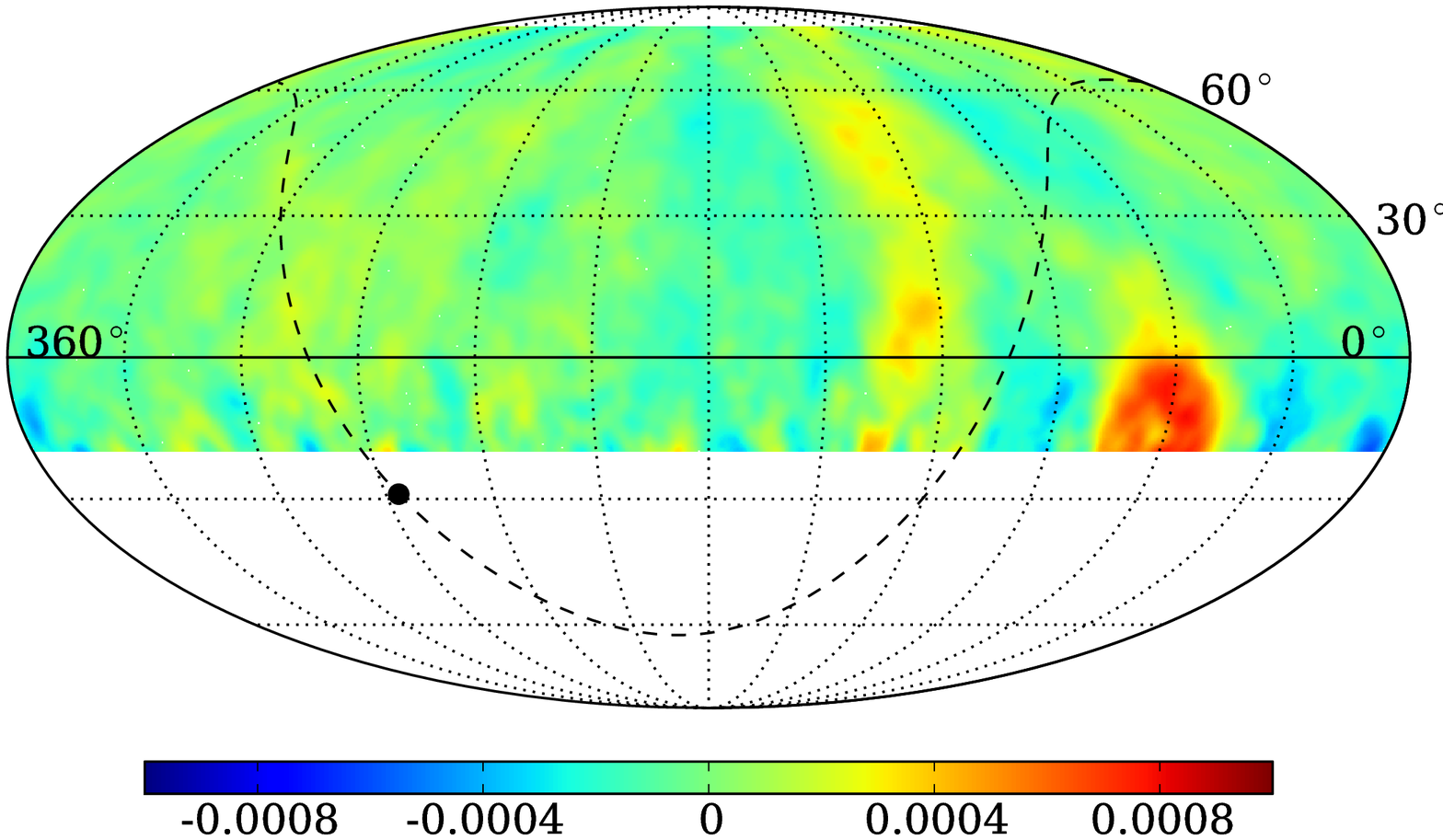}} \\
(a)\\
{\includegraphics[angle=0,width=.4\textwidth]{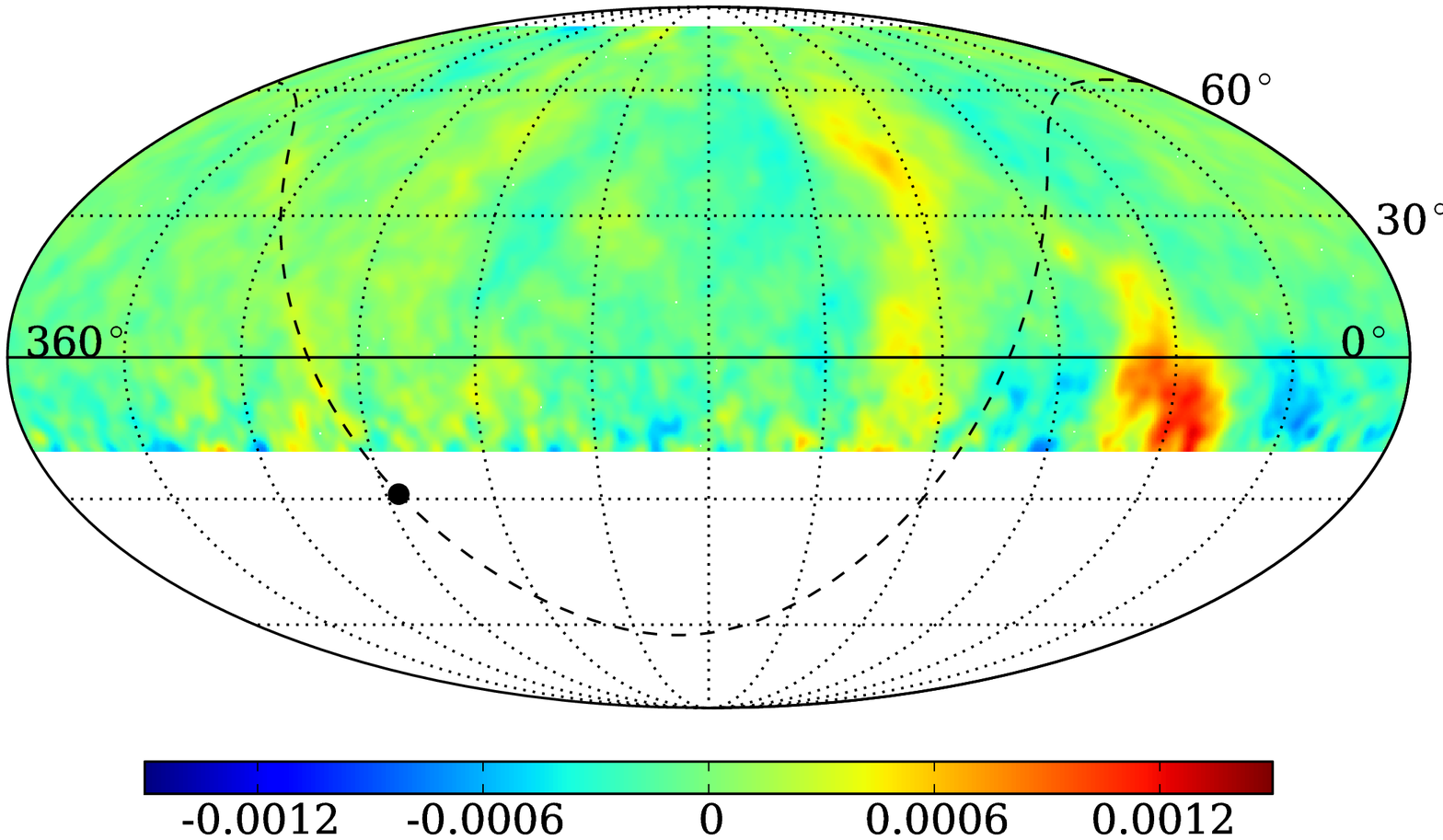}}\\
(b)\\
{\includegraphics[angle=0,width=.4\textwidth]{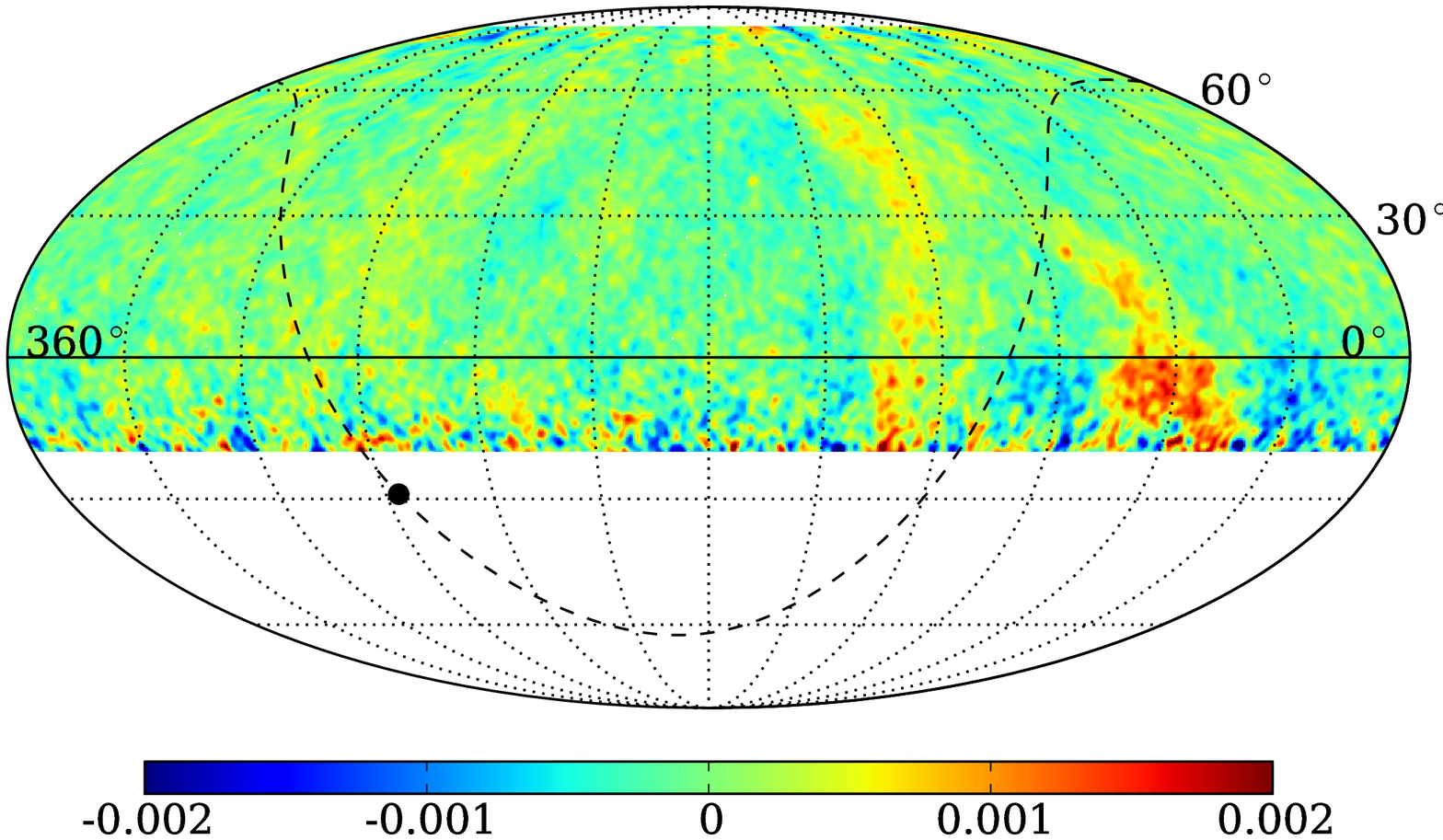}} \\
(c)\\
{\includegraphics[angle=0,width=.4\textwidth]{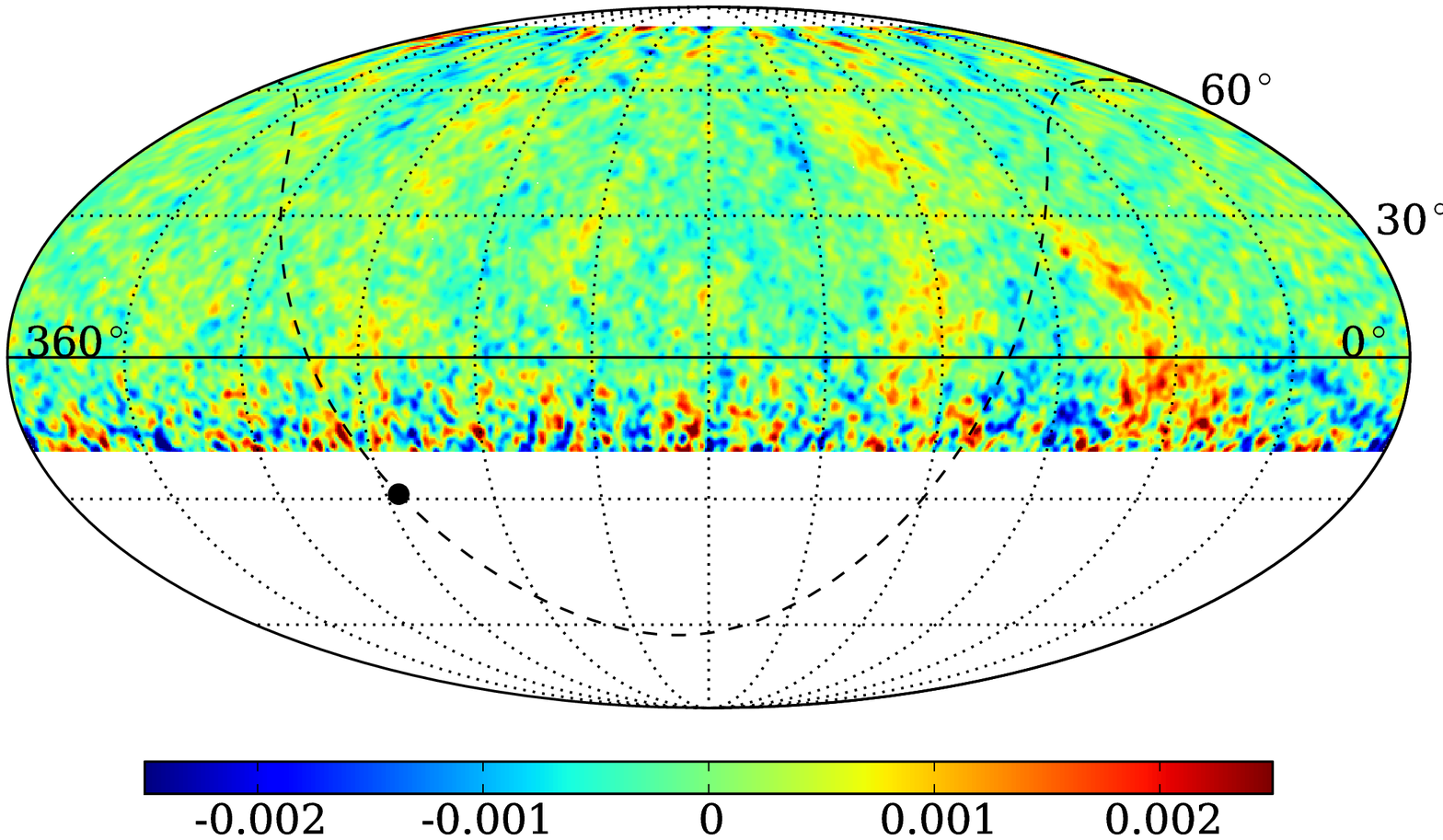}}\\
(d)
\caption{The fractional CR excess with respect to the estimated background is shown for different shower multiplicities: (a) 25 - 39, (b) 40 - 99, (c) 100 - 249, (d) 250 - 629 fired strips on the ARGO-130 central carpet. \emph{Mollweide projections.}}
\label{fig:msa-energy}
\end{figure}

The Fig. \ref{fig:msa-energy} shows the dependence of the MSA on the multiplicity of the detected showers. The color scale is symmetric and has been optimized for each map, so that the intensity range spans the full color spectrum.
%
%

The evolution of the medium-scale anisotropy can be appreciated by examining the multiplicity sequence. The regions 1 and 2 are well visible up to the 250-630 multiplicity range, the regions 3 and 4 are noticeable mostly in the range 40-250.

If the evolution of the region 1 is followed, it can be noticed that up to multiplicities as high as 100, the sub-region 1L is much brighter than the small spot 1U. For greater multiplicities, the region 1U intensity becomes as high as for 1L, even the sub-region 1U emits more than any other region in the map, whereas the intensity of the sub-region 1L decreases. Because of poor statistics, fluctuations begin to disturb the imaging from 250 hits on, that is why the interval 250-630 is the last to be represented (Fig.  \ref{fig:msa-energy}(d) ).
On the other hand, if the evolution of the region 2 is considered, it can be seen that the lower spot (sub-region 2L) considerably fades off from 100 hits on. 
%
\bfi{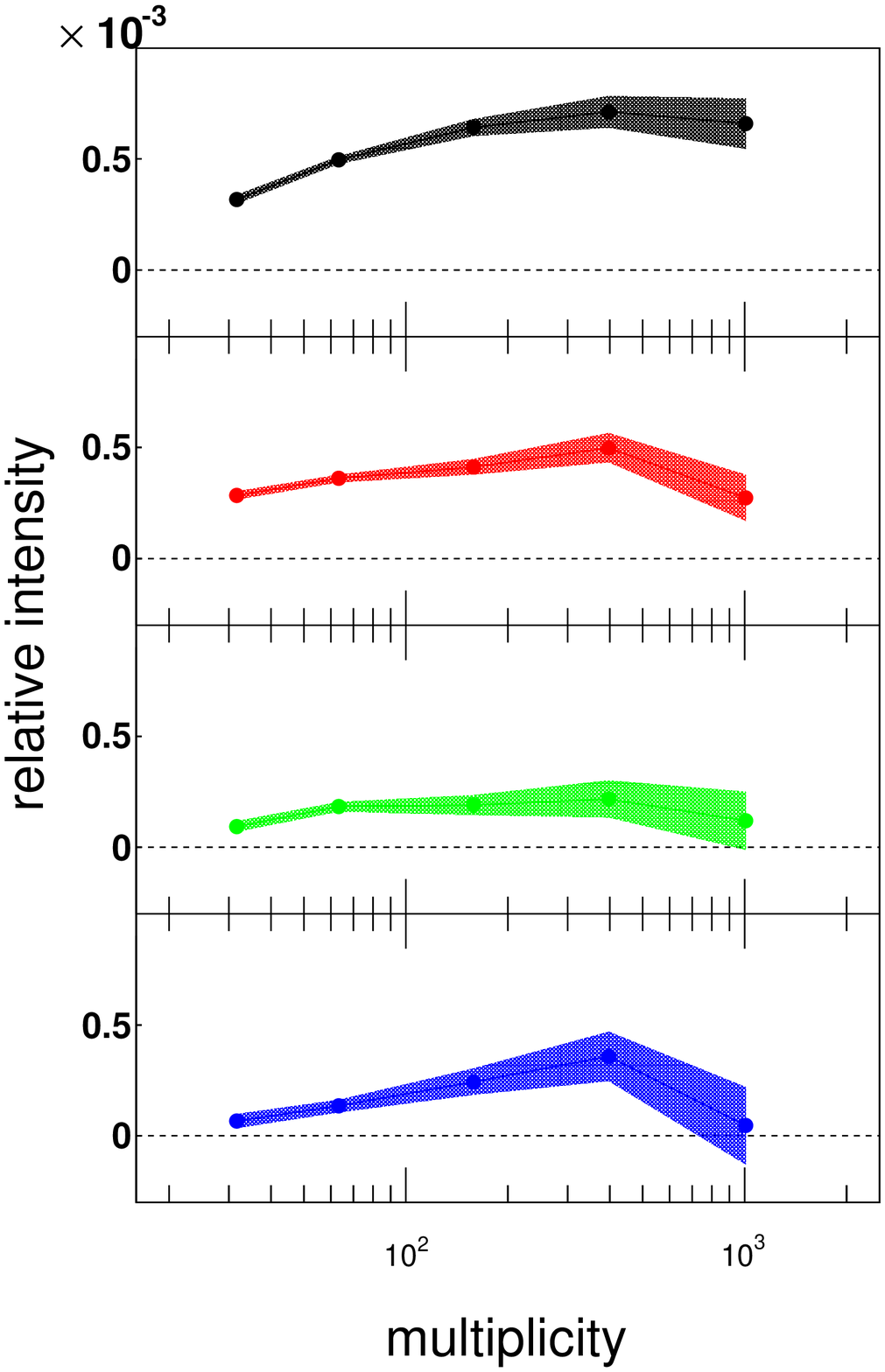}\\
  \caption{Size spectrum of the four MSA regions observed by ARGO-YBJ (regions 1 to 4 starting from the top). The vertical axis represents the relative excess $(e-b)/b$. 
  The statistical errors are represented as coloured bands around the experimental points.}
  \label{fig:energysp}
\efi
%

The Fig. \ref{fig:energysp} reports the multiplicity spectra for the anisotropy regions 1-4 (top-down). The number of events collected within each region are computed for the event map $e$ as well as for the background one $b$. The relative excess $(e-b)/b$ is computed for each multiplicity interval. The horizontal axis reports the multiplicity, the vertical one the relative intensity. 

The black plot reports the region 1 multiplicity spectrum. It is the hardest one detected by \argo and it shows a flattening around multiplicity 400 at relative intensity $\sim0.7\times10^{-3}$.
The region-2 multiplicity spectrum (red plot) is flatter than the one of region 1 and it turns out to be compatible with the constant result obtained by Milagro \cite{milagro2008}. The average intensity is $\sim0.35\times10^{-3}$.
Similar results are obtained for the region 3 (green graph), although the intensity is settled around $\sim0.2\times10^{-3}$.
The region 4 (blue graph), the least significant one, has a steep spectrum which rises up at a multiplicity between 300 and 400. 

Looking at the width of the error band (statistical error), it appears that for each region the multiplicity analysis gives a significant result up to $N=300-400$. Instead, the high-multiplicity measurements are significant only for the regions 1 and 2, as the region 3 and 4 average excess is compatible with a null result.

The emission from region 1 is so intense and its observation so significant that interesting information can be obtained from the analysis of the multiplicity-energy relation in the sub-regions of its parametrization. In fact, the comparison of sub-region spectra is an important tool to check whether sub-regions are just geometrical parameterizations of the observed anisotropies, or they host different sources with various emission mechanisms.
%
\begin{figure}[!htbp]
  \centering
  {\includegraphics[width=0.45\textwidth]{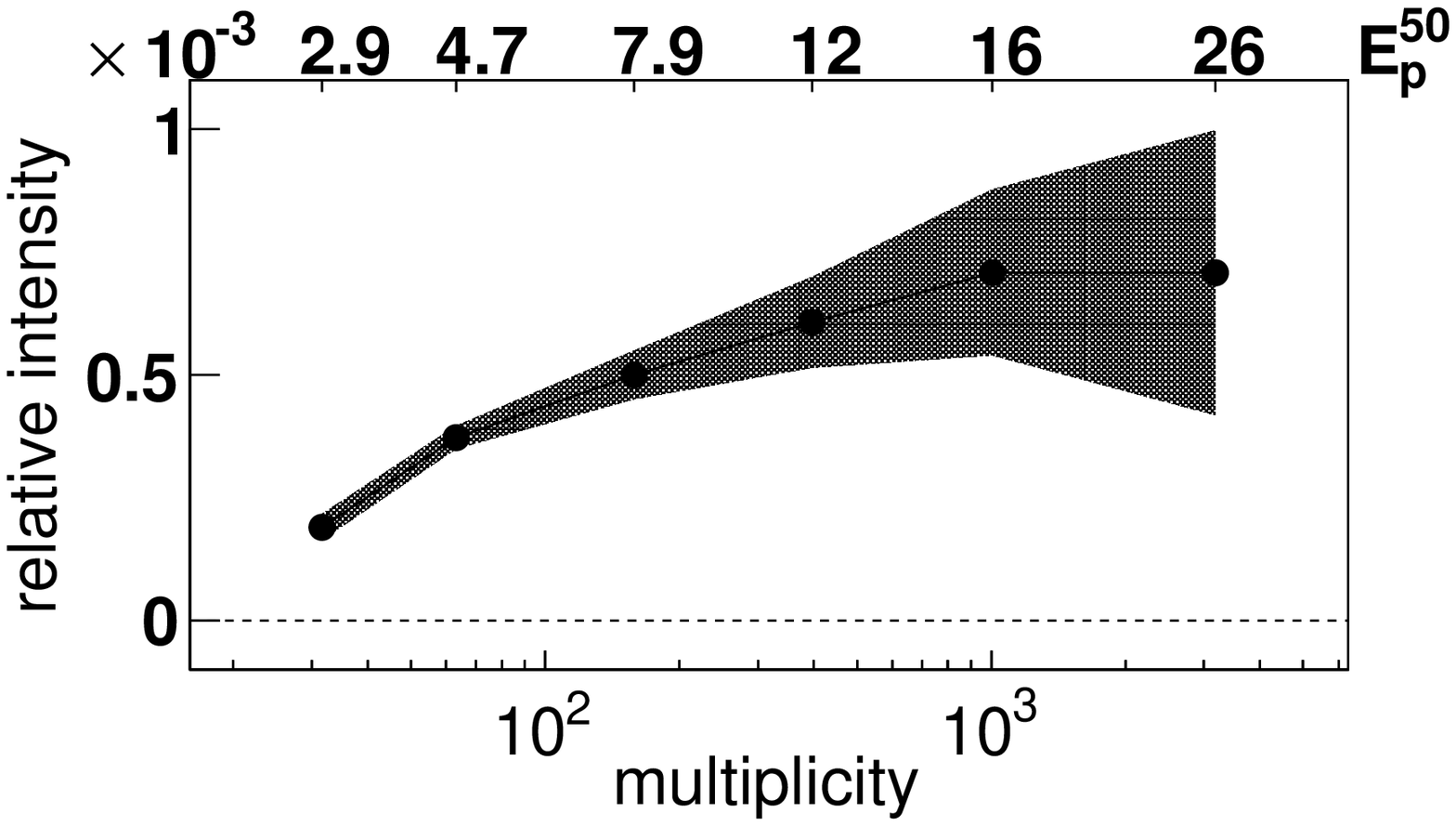}}\\(a)\\
  {\includegraphics[width=0.45\textwidth]{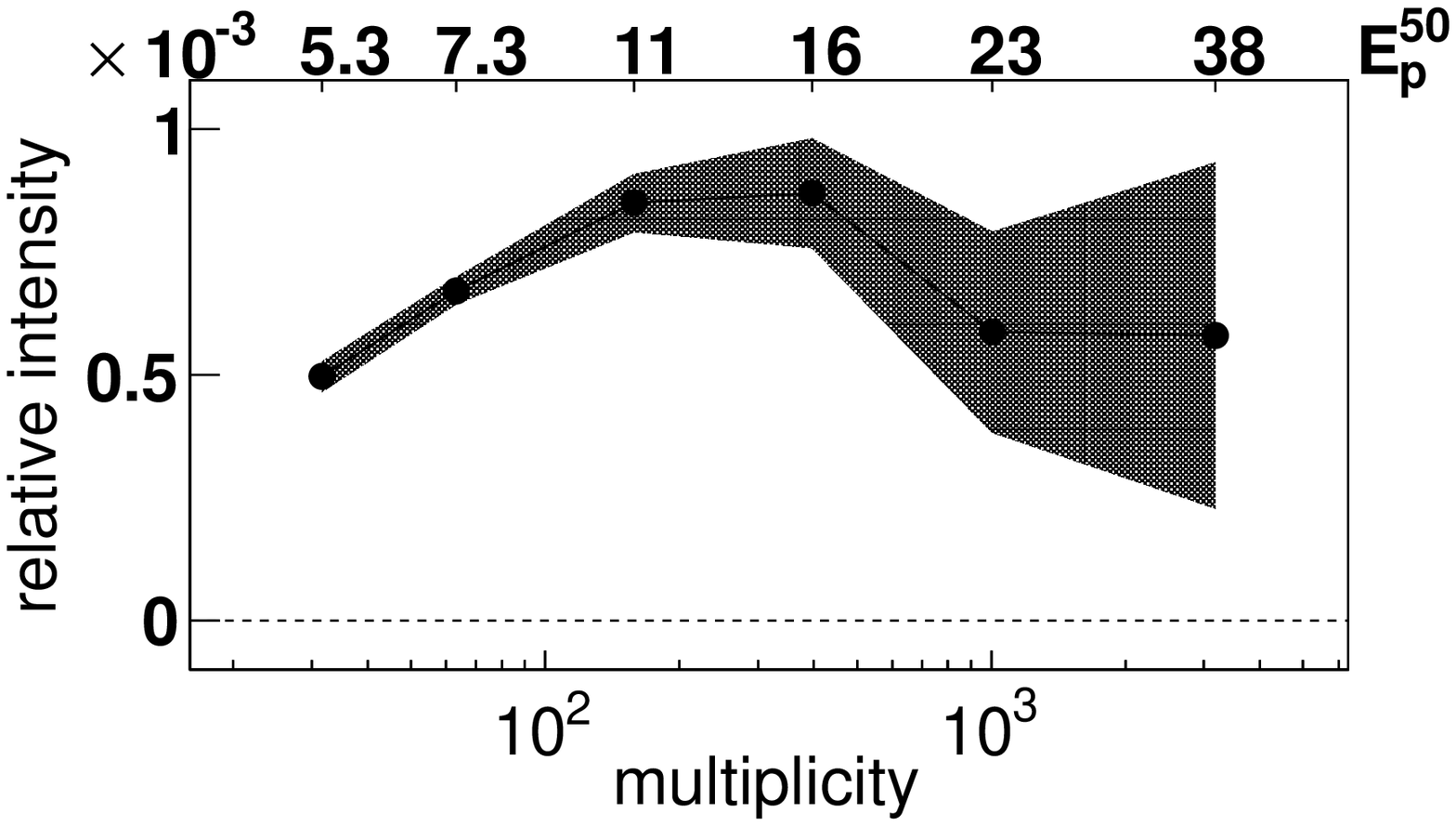}}\\(b)
  \caption{Multiplicity spectra of the sub-regions 1U (a) and 1L (b). The vertical axis represents the relative excess $(e-b)/b$. The upper horizontal scale shows the corresponding proton median energy (TeV). Six multiplicity intervals were used instead of the five described in section \ref{sec:datanalysis}, see text for details.}
  \label{fig:energysp-1}
\end{figure}

The Fig. \ref{fig:energysp-1} poses the spectrum of the sub-regions 1U and 1L, with energy scales computed for a proton point-source having the average declination of each sub-region. 
To get more refined results at high energy, the last multiplicity bin (more than 630 fired strips) was split into $630-1599$ and $\geq 1600$.
For the region 1L a cut-off around 15-20 TeV can be noticed. The statistics at high multiplicity is very poor and does not allow to establish whether the cut-off continues at higher energy or not. Conversely, for region 1U a constantly increasing trend is obtained up to 26 TeV, what marks a possible difference between the sub-regions. Such a result has to be interpreted in the framework of a declination-dependent energy response, to ascertain if a cut-off is present at higher energy. Within the error bars it would be compatible with findings about the region ``A'' by Milagro \cite{milagro2008}.

 As already said, the elemental composition and the energy spectrum are not known and that of CR protons is just an hypothesis. The ``photon'' hypothesis cannot be excluded \emph{a priori}, because in this work no gamma/hadron discrimination algorithms are applied. Even regions 1 and 2 exceed so much the Milagro parametrization that the conclusion about regions A and B not due to photons cannot be drawn.

Concerning the sub-parts of regions 2, 3 and 4, no significant features were found in their energy spectra, thus there is no reason to consider them more than just a simple geometrical parameterization.

%
\subsection{Dependence on time}

The Milagro collaboration found evidence that in their regions ``A'' and ``B'' the fractional excess was lower in the summer and higher in the winter \cite{milagro2008}.
We investigated the stability of the fractional excess in all four regions with data recorded by ARGO-YBJ in the 2007 - 2012 years. 
As it can be seen in Fig. \ref{fig:time-stab}, there is no evidence either of a seasonal variation or of constant increasing or decreasing trend of the emission, as expected from the cancellation of many systematics in measuring relative quantities. The average flux values are $(0.50\pm0.04)\,10^{-4}$, $(0.37\pm0.03)\,10^{-4}$, $(0.16\pm0.03)\,10^{-4}$ and $(0.14\pm0.03)\,10^{-4}$ for regions 1, 2, 3 and 4 respectively ($\chi^2/\textrm{d.o.f.}$ 23/18, 33/18, 38/18 and 28/18).

%
\bfi{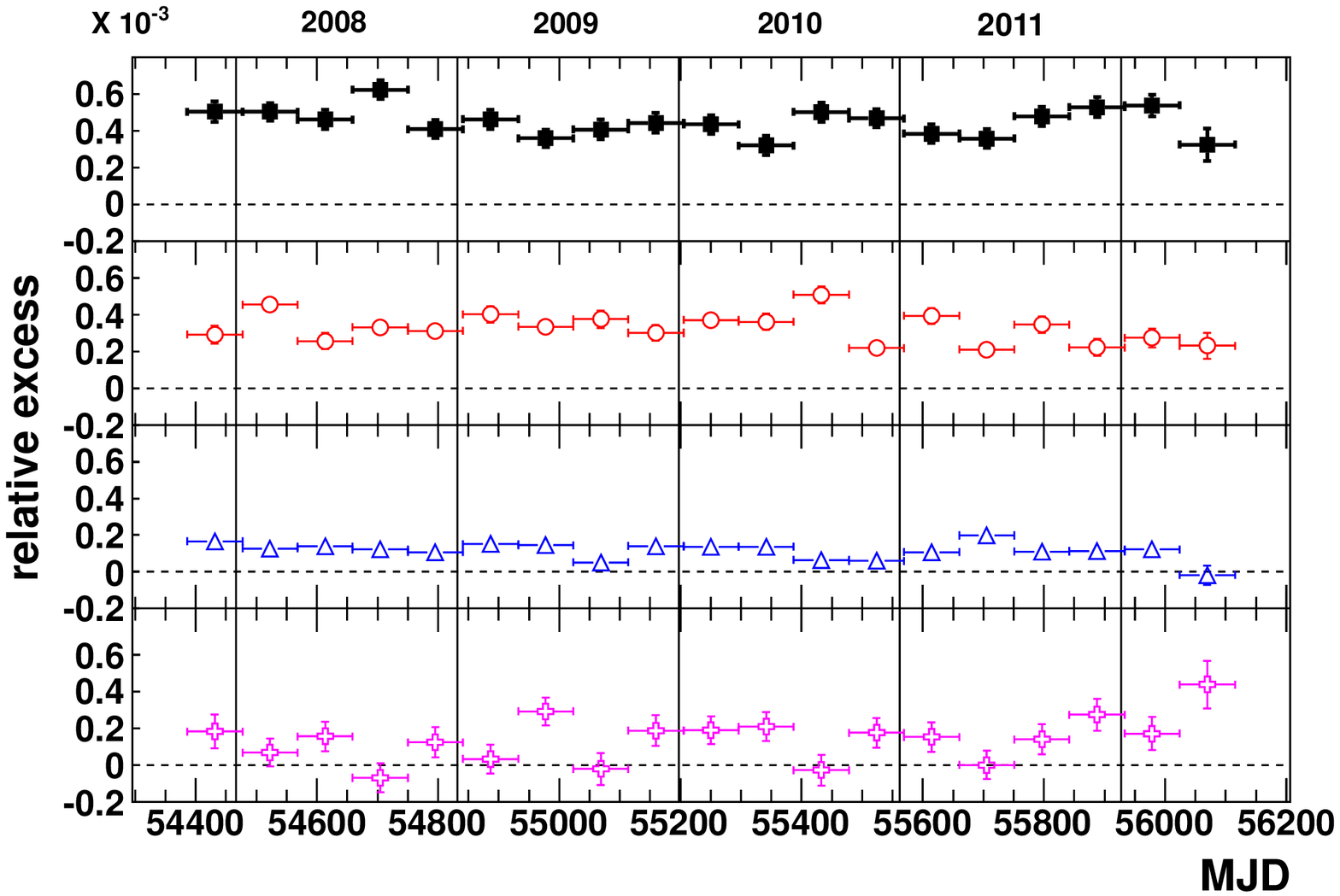}\\
  \caption{The relative event excess with respect to the background in the regions 1 to 4 as a function of the observation time is shown starting from the top. The plots refer to events with a multiplicity N$_{strip}>$25. The time-bin width is approximately 3 months.}
  \label{fig:time-stab}
\efi
%

\section{Conclusions}

In the last years the CR anisotropy came back to the attention of the scientific community, thanks to several new two-dimensional representations of the CR arrival direction distribution. 

Some experiments collected so large statistics as to allow the investigation of anisotropic structures on smaller angular scale than the ones corresponding to the dipole and the quadrupole.
Along this line, the observation of some regions of excess as wide as $\sim$30$^{\circ}$ in the rigidity region $\sim$1 - 20 TV stands out. The importance of this observation lies in the unexpected confinement of a large flux of low-rigidity particles in beams which are too narrow to be accounted for by the local magnetic-field bending power.

In this paper we reported the ARGO-YBJ observation of anisotropic structures on a medium angular scale as wide as $\sim$10$^{\circ}$ - 45$^{\circ}$, in the energy range 10$^{12}$ -10$^{13}$ eV. The intensity spans from 10$^{-4}$ to 10$^{-3}$, depending on the selected energy interval and sky region.

For the first time the observation of new MSA structures throughout the right ascension region $195^{\circ}-290^{\circ}$ is reported with a statistical significance above 5 s.d.
The size spectra of the detected excess regions look quite harder than the corresponding ones for the CR isotropic flux and a cut-off around 15-20 TeV is observed for those regions where the dynamics of the experiment is sufficiently extended.

As discussed in detail in the Introduction, although some hypotheses have been made, models to explain the whole set of observations are missing and deep implications on the physics of CRs in the Local Interstellar Medium are expected. 

Given the relevance of the subject and the uncertainty its elemental composition, a joint analysis of concurrent data recorded by different experiments in both hemispheres, as well as a correlation with other observables like the interstellar energetic neutral-atom distribution \cite{ibex09,ibex11}, should be a high priority to clarify the observations.
Further studies of the \argo collaboration are in progress in order to achieve a better separation of the signal in the harmonic space, as well as to investigate the nature of the phenomenon.

\begin{acknowledgments}
This work is supported in China by NSFC (No. 10120130794), the Chinese Ministry of Science and Technology, the Chinese Academy of Sciences, the Key Laboratory of Particle Astrophysics, CAS, and in Italy by the Istituto Nazionale di Fisica Nucleare (INFN). We also acknowledge the essential support of W. Y. Chen, G. Yang, X. F. Yuan, C. Y. Zhao, R. Assiro, B. Biondo, S. Bricola, F. Budano, A. Corvaglia, B. D'Aquino, R. Esposito, A. Innocente, A. Mangano, E. Pastori, C. Pinto, E. Reali, F. Taurino, and A. Zerbini, in the installation, debugging, and maintenance of the detector.
\end{acknowledgments}


\end{document}